\documentclass[aps,twocolumn,nofootinbib,preprintnumbers,superscriptaddress]{revtex4}

\usepackage{color}
\usepackage{tikz}

\newcommand{\nb}[1]{\color{blue}}

\newcommand{\hl}[1]{\color{magenta}}

\usepackage[
	  pagebackref=false,
	  colorlinks=true,
      linkcolor=blue,
      urlcolor=blue,
      filecolor=black,
      citecolor=red,
      pdfstartview=FitV,
      pdftitle={},
        pdfauthor={},
        pdfsubject={},
        pdfkeywords={},
        pdfpagemode=None,
        bookmarksopen=true
      ]{hyperref}

\usepackage[normalem]{ulem}
\usepackage{amsmath}
\usepackage{enumerate}
\usepackage{amsfonts}
\usepackage{epsfig}
\usepackage{mathbbol}

\newcommand\constructosum[3]{%
    \begin{tikzpicture}[baseline=(char.base), inner sep=0, outer sep=0]
        \draw (#1,0) circle (#2);
        \node (char) at (0,0) {$#3\sum$}; 
    \end{tikzpicture}%
}

\newcommand{\modtwosum}{\mathop{\mathchoice
        {\constructosum{-0.3ex}{0.1}{\displaystyle}}
        {\constructosum{-0.3ex}{0.06}{\textstyle}}
        {\constructosum{-0.2ex}{0.04}{\scriptstyle}}
        {\constructosum{-0.15ex}{0.03}{\scriptscriptstyle}}
    }\displaylimits
}

\def\Tr{\mathop{\rm Tr}}
\def\tr{\mathop{\rm tr}}

\newcommand\p{\ensuremath{\partial}}

\newcommand{\be}{\begin{equation}}
\newcommand{\ee}{\end{equation}}
\newcommand{\bea}{\begin{eqnarray}}
\newcommand{\eea}{\end{eqnarray}}
\newcommand{\bega}{\begin{gather}}
\newcommand{\eega}{\end{gather}}

\newcommand{\bi}{\begin{itemize}}
\newcommand{\ei}{\end{itemize}}
\newcommand{\ben}{\begin{enumerate}}
\newcommand{\een}{\end{enumerate}}
\newcommand{\bca}{\begin{cases}}
\newcommand{\eca}{\end{cases}}
\newcommand{\bln}{\begin{align}}
\newcommand{\eln}{\end{align}}
\newcommand{\bst}{\begin{split}}
\newcommand{\est}{\end{split}}
\def\ie{\begin{equation}\begin{aligned}}
\def\fe{\end{aligned}\end{equation}}
\newcommand{\bma}{\le(\begin{matrix}}
\newcommand{\ema}{\end{matrix}\ri)}

\newcommand\vep{\varepsilon}

\def\le{\left}
\def\ri{\right}

\begin{document}

\title{\textbf{Effective response theory
    for Floquet topological systems}}

\author{Paolo Glorioso}
\affiliation{Kadanoff Center for Theoretical Physics, University of Chicago,
Chicago, Illinois 60637, USA}
\author{Andrey Gromov}
\affiliation{Department of Physics, Brown University, 182 Hope Street, Providence, RI 02912, USA}
\author{Shinsei Ryu}
\affiliation{Kadanoff Center for Theoretical Physics, University of Chicago,
Chicago, Illinois 60637, USA}
\affiliation{James Franck Institute, University of Chicago, Chicago, Illinois 60637, USA}

\begin{abstract}
 We present an effective field theory approach to the topological response of Floquet systems with symmetry group $G$.
  This is achieved by introducing a background $G$ gauge field in the Schwinger-Keldysh formalism, which is suitable for far from equilibrium systems. We carry out this program for chiral topological Floquet systems (anomalous  Floquet-Anderson insulators) in two spatial dimensions,
  and the group cohomology models of topological Floquet unitaries.
  These response actions serve as many-body topological invariants for
  topological Floquet unitaries. The effective action approach also leads us to
  propose
  novel topological response functions.
\end{abstract}

\maketitle
\tableofcontents

\newpage

Topological phenomena in periodically driven systems (Floquet systems) have been widely discussed recently.
For recent review articles, see, e.g., \cite{2017RvMP...89a1004E, 2018arXiv180403212O, 2019arXiv190501317H}.
In a typical setup, we consider dynamics governed by
the Hamiltonian which depends periodically on time $t$:
${H}(t + T) = {H}(t)$.
Correspondingly, we consider the time-evolution operator
\begin{align}\label{evol}
  {U}(t,t_0) = \mathrm{T} \exp
  \left[-i \int^{t}_{t_0} dt' {H}(t') \right]\ ,
\end{align}
where $\mathrm{T}$ represents time-ordering.
As a slight variation of the problem,
we also consider a periodic time evolution described by a periodic unitary
${U}(t+T)={U}(t)$, without mentioning Hamiltonians.

It has been discovered that such periodic drive can give rise
to topological phenomena of at least two different kinds:
(i) The periodic drive can turn a non-topological static system 
into a topological system, which can essentially be understood as a static topological system.
(ii) The periodic drive can give rise to a topological phenomenon, which is
unique to periodically driven systems, and has no analogue in static systems.
Initial studies of topological Floquet systems were limited
to the first kind of dynamical topological phenomena
\cite{2009PhRvB..79h1406O,2010PhRvB..82w5114K,
  2010PhRvL.105a7401I, 2011NatPh...7..490L, 2011PhRvB..84w5108K}.
On the other hand, phenomena of the second kind have been discovered and studied more recently
\cite{2011PhRvL.106v0402J,
  2015PhRvL.114j6806C,
  2016PhRvB..94l5105R,
  2016PhRvB..94h5112V,
  2016PhRvB..93x5145V,
  2016PhRvB..93x5146V,
  2016PhRvB..93t1103E,
  2016PhRvX...6d1001P,
  2018arXiv181209183G,
  2017PhRvB..95o5126P,
  2017PhRvB..95s5128R}.
Of particular interest in this paper are
topological chiral Floquet drives (anomalous Floquet-Anderson insulators) in two spatial dimensions
\cite{
  2012arXiv1212.3324R,
  2016PhRvX...6b1013T,
  2017PhRvL.119r6801N,
  2016PhRvX...6d1070P,
  2017arXiv170307360F,
  2017PhRvB..96o5118R,
  2017PhRvB..95s5155M},
which are characterized by the 3d winding number of their single-particle Floquet
unitary operators.
%
%

The purpose of this paper is to develop an effective
response field theory approach to Floquet topological systems.
Our primary focus will be topological phenomena
of the second kind
which are intrinsic to the non-equilibrium nature of periodically driven systems.

For ``static'' topological phases of matter, their descriptions in terms of
effective response field theories have been well developed
(see, for example, \cite{2008PhRvB..78s5424Q}).
A canonical example is the Chern-Simons effective field
theory describing the response of quantum Hall states in (2+1)d.
One first introduces suitable background gauge fields; for the case of particle number
conserving systems, we can introduce
the background $U(1)$ gauge field $A$.
We can then integrate out the dynamical ``matter'' fields:
\begin{align}
  Z [A]                         %
  &=
    \mathrm{Tr}\,
    \left[
    \mathrm{T}_{\tau} e^{ - \int^{\beta}_0 d\tau H(\tau; A) }
    e^{ - \int^{\beta}_0 d\tau \int d^dr\, i A_0 j^0}
    \right]
    \nonumber \\
  &=
  \int \mathcal{D}[\psi^{\dag},\psi]
  \exp
  \left(-S[\psi^{\dag},\psi, A] \right)
          \nonumber \\
  &\equiv
  \exp
  \left(- S_{{\rm eff}}[A] \right)\ ,
  \label{path integral}
\end{align}
where we are working with Euclidean signature.
For integer quantum Hall systems, it is known that the topological part of
the effective action is purely imaginary and given by the Chern-Simons term
\begin{align}
  \label{CS}
  S_{{\rm eff}}[A]
  &=
  \frac{ i \nu}{4\pi} \int A\wedge dA
    \nonumber \\
  &=
  \frac{ i \nu}{4\pi} \int d \tau d^2r\,
  \varepsilon^{\mu\nu\lambda}
  A_{\mu} \partial_{\nu} A_{\lambda},
\end{align}
where $\nu$ is an integer.
The modulus of the (topological part of the) partition function
is independent of $A$,
and can be normalized to be 1, $|Z[A]|\simeq 1$.

Topological effective response field theories describe the properties of quantum many-body systems which are stable against interactions.
It should be also emphasized that
starting from effective response field theories
it is often possible to construct
explicit formulas for
many-body topological invariants.
(See, for example,
\cite{2017PhRvB..95t5139S,2018PhRvB..98c5151S}.)
In deriving the effective field theory,
it is crucial that we deal with
gapped (topological) phases, where matter fields represent ``fast'' degrees of freedom and can then be ``safely'' integrated over, i.e.,
the integration over the matter field can be controlled
by the inverse gap expansion and leads to
a local effective action.
It should also be noted that in the presence of the gap,
the topological term in the response field theory
encodes purely the properties of the ground state.
In other words, the presence/absence of the topological term
in the response theory can be deduced from
the adiabatic response of the ground state,
without discussing gapped excited states.
For example,
in the case of the integer quantum Hall effect,
the coefficient of the Chern-Simons term is
expressed in terms of the many-body Chern number.

In this paper, we develop an effective response theory
approach for periodically driven (topological) systems,
paralleling effective response theories for static
topological phases with symmetry.
Specifically,
we will work with the Schwinger-Keldysh generating functional of Floquet unitaries,
\begin{align}
  Z[A_1, A_2] =
  \mathrm{Tr}
  \left[
  U( A_1) \rho_0 U^{\dag}( A_2)
  \right]\ ,
\end{align}
where we have introduced external (non-dynamical) $U(1)$ gauge fields $A_{1}$
and $A_2$, which couple to the evolution operator $U(t)$ and to its conjugate
$U^\dag(t)$, respectively\footnote{
  One can also promote static background gauge fields to dynamical ones,
  by integrating over the gauge field. The effect of such dynamical gauging
  was discussed in \cite{2017PhRvB..95o5126P}.
  In this work, we will confine ourselves to background gauge fields.
}.
The initial state $\rho_0$, will be taken to be
a Gibbs ensemble at infinite temperature, $\rho_0 \sim e^{\alpha Q}$,
where $\alpha$ is a chemical potential, and $Q$ is the $U(1)$ charge operator.
We will demonstrate that
the Schwinger-Keldysh generating functional
$W[A_1,A_2] = -i \log Z[A_1, A_2]$
for many-body localized Floquet systems
is a local functional of $A_1$ and $A_2$,
and, furthermore, encodes the topology of the system. In fact, we will see that this framework can be applied to many-body localized systems whose explicit time dependence is not necessarily periodic, as the topological origin of our response functional --i.e., its independence on smooth deformations of the system-- is completely unrelated to time periodicity of the microscopic system. Periodicity, combined with other properties, will only be used to show quantization of topological response.

The paper is organized as follows. In Sec.\ \ref{Schwinger-Keldysh response},
we explain how to apply the Schwinger-Keldysh approach to topological periodically driven systems.
In
Sec.\ \ref{Chiral Floquet drive},
we consider
chiral Floquet drives (topological Floquet Anderson insulators)
in two spatial dimensions,
for which we explicitly compute the
Schwinger-Keldysh effective action,
and identify the topological term.
In Sec. \ref{sec:eft}, we describe two generalizations of effective response
corresponding to candidate
novel topology which was not discussed before.
Further material and technical details are discussed in the Appendix.
In Appendix \ref{Group cohomology models},
we study yet another class of topological Floquet drives,
those which are constructed by using the group cohomology.
There, we find that the topological terms of
the Schwinger-Keldysh functional
are members of (labeled by) $H^d(G, U(1))$ where
$G$ is the symmetry group and $d$ is the spatial dimension.
In Appendix \ref{Channel-state map approach},
we discuss an approach based on
the so-called channel-state map,
which provides a perspective complementary to the Schwinger-Keldysh approach.

%

\section{Schwinger-Keldysh response}
\label{Schwinger-Keldysh response}

\subsection{Generalities}
In this section, we introduce the basic framework that will be used as a
systematic approach to topological Floquet phases.
While our interest lies in Floquet systems,
we shall start with general discussions that can be
applied to any time-dependent Hamiltonian $H(t)$. A modern introduction to the Schwinger-Keldysh formalism can be found in \cite{kamenev_2011,2018arXiv180509331G}.

We will assume that $H(t)$ possesses a $U(1)$ symmetry, and we couple it to an external gauge field $A_\mu(t,\vec r)$, so that the evolution operator is given by
\begin{align}
  {U}(t_1,t_0;A) = \mathrm{T} \exp
  \left[-i \int^{t_1}_{t_0} dt' {H}(t';A) \right]\ ,
\end{align}
where $H(t;A)$ is the Hamiltonian coupled to $A_\mu(t,\vec{r})$. The current conjugate to $A_{\mu}$ will be denoted as $J^{\mu}$.

We introduce the Schwinger-Keldysh generating functionals
$Z[ A_1, A_2]$
and $W[A_1,A_2]$ by
\cite{kamenev_2011,2018arXiv180509331G}
\begin{align}
  \label{genf}
  Z[A_1,A_2]
  &= e^{ i W[A_1,A_2]}
    \\
  &=
  \mathrm{Tr}
  \left[
  U(t_1,t_0; A_1) \rho_0 U^{\dag}(t_1,t_0; A_2)
  \right]\ ,
    \nonumber
\end{align}
where $\rho_0$ is the initial state of the system at $t=t_0$.
The operator inside the trace can be thought of as the time evolution of the
density matrix $\rho_0$,
$\rho(t_1)=U(t_1,t_0;A_1)\rho_0U^\dag(t_1,t_0;A_2)$, where
each factor of the evolution is coupled to a different gauge field
$A_{1\mu}$ and $A_{2\mu}$.

In typical applications,
we adiabatically switch on perturbations causing
nonequilibrium dynamics. It is then convenient to start the time-evolution
from $t_0= -\infty$
with the initial state $\rho_0$ in the remote past
which is chosen as an equilibrium state.
We also send $t_1\to +\infty$ when discussing correlation functions
with operators located at arbitrary late times.
Then, the Schwinger-Keldysh contour runs from $-\infty$ to $+\infty$ and back.
One striking feature is that this approach does not require knowing the final state

The Schwinger-Keldysh trace with background \eqref{genf} provides a compact and efficient way to encode various non-equilibrium correlation functions. Indeed, differentiating $Z[A_1,A_2]$ $n$ times with respect to $A_{1\mu}$ and $m$ times with respect to $A_{2\mu}$ leads to a correlation function of $n$ time ordered and $m$ anti-time ordered currents $J^\mu$
\begin{align}\label{tor}
  &
  \mathrm{Tr}\left[\rho_0
    \mathrm{T}(J^\mu(x_1)\cdots)
  \tilde{\mathrm{T}}(J^\alpha(x_{n+1})\cdots )\right]
    \\
  &\quad
=\left.\frac 1{i^n(-i)^m}\frac{\delta^{n+m} e^{iW[A_1, A_2]}}{\delta A_{1\mu}(x_1)\cdots\delta
    A_{2\alpha}(x_{n+1})\cdots}\right|_{A_{1,2}=0}\ ,
  \nonumber
\end{align}
where $x=(t,\vec{r})$ and
$\tilde{\mathrm{T}}$ represents anti-time ordering.

The generating functional $W[A_1,A_2]$ should satisfy certain basic properties due to unitarity of the evolution:
\be
\begin{gathered}\label{cons}
  W[A_1,A_2]=-W^*[A_2,A_1],\quad W[A,A]=0,
  \\
  \text{Im}\, W[A_1,A_2]\geq 0
\end{gathered}
\ee
where the first two can be seen straightforwardly from the definition (\ref{genf}),
while the last condition follows from the fact
that the absolute value of the trace of the operator
$U(\infty,-\infty; A_1) \rho_0 U^{\dag}(\infty,-\infty; A_2)$
is bounded by unity
\cite{2016arXiv161207705G,2018arXiv180509331G}.

%

\subsection{Application to (topological) Floquet systems}

We will now apply
the Schwinger-Keldysh formalism
to study topological properties of Floquet systems.

\paragraph{Choice of the initial state}
For static systems, one typically chooses the initial state $\rho_0$ to be the
ground state or the thermal state.
In the case of our interest,
we observe that the time dependence of the Hamiltonian is not slow compared to
the energy gap of the instantaneous Hamiltonian $H(t)$, and thus there is no notion of ground state, nor of thermal
equilibrium.
The most natural choice in this context,
in the absence of any symmetry,
is to choose $\rho_0$ to be
the infinite temperature state,
\begin{align}
  \label{inf T state}
  \rho_0 = I/\mathcal{N}\ ,
\end{align}
where $I$ is the identity operator and the normalization factor $\mathcal{N}$
is the dimension of the Hilbert space. In the Appendix \ref{Channel-state map approach}, we will see that with this choice of
initial state, the Schwinger-Keldysh trace can be viewed
as an inner product of unitaries when unitaries are mapped to
states by using the channel-state map
(the so-called Choi-Jamio\l kowski isomorphism).

In the presence of a symmetry,
the most natural choice of $\rho_0$ is
the Gibbs ensemble formed by the conserved charges of the system,
\begin{align}
  \label{rho}
  \rho_0=\frac{e^{\alpha Q}}{\Tr e^{\alpha Q}}\ ,
\end{align}
where $Q$ is the charge operator (number operator) associated to the $U(1)$
symmetry, and
the parameter $\alpha$ plays the role of chemical potential.
Instead of introducing $\alpha$ in the initial state $\rho_0$,
$\alpha$ can also be introduced as the difference between
the (uniform and time-independent)
temporal component of $A_1$ and $A_2$ in
$U(t_1,t_0;A_1)$ and $U(t_1,t_0;A_2)$.

This choice of initial state allows us to put our focus on properties of evolution operators themselves,
rather than the time evolution of individual states.
(See, for example, \cite{2018arXiv180408638R} which also uses the infinite temperature state.)
We also recall that under Floquet time evolution,
states may indefinitely be heated up by the drive, which may wash out
any topological phenomena. Various mechanisms in the literature are used to prevent this (e.g., many-body localization
or prethermalization
\cite{2015PhRvL.115y6803A,
2017CMaPh.354..809A,
2016PhRvL.116l0401M,
2016AnPhy.367...96K,
2017PhRvX...7a1026E}).
It is also worth recalling that eigenstates of Floquet
unitaries are all expected to behave similarly,
e.g., no mobility gap separating ergodic and many-body localized states.

\paragraph{Choice of the Schwinger-Keldysh contour}

We now describe our choice of the Schwinger-Keldysh contour.
First, we note that there are characteristic values of times,
integer multiples of
the period of the Floquet drive $T$.
In our discussion,
we will evaluate the Schwinger-Keldysh generating functional
for $t_1-t_0 = ({\rm integer}) \times T$.
At these values the generating functional will exhibit additional important
properties in relation to topology when dealing with special models
-- see Sec.\ \ref{Chiral Floquet drive}.

Second, for generic systems, it will be important to take the integers $m,n$ to be large.
A convenient object to study response to $A_\mu$ is then
\begin{align}
  \begin{gathered}
    Z[A_1, A_2]=
    e^{iW[A_1, A_2]}=
    \\
  \lim_{ {\kappa}\to \infty}
  \mathrm{Tr}
  \left[
  U({\kappa}T,-{\kappa}T; A_1) \rho_0 U^{\dag}({\kappa}T,-{\kappa}T; A_2)
  \right]\ ,
  \end{gathered}\label{zaa}
\end{align}
where
we chose $t_1=-t_0 = \kappa T$ with ${\kappa}$ a half-integer;
for simplicity,
we have chosen the Schwinger-Keldysh contour to be symmetric
around $t=0$.
As we will see, the infinite time limit guarantees that, for generic models, when the system
is in the localized regime, only topological contributions will survive, as in the infinite time limit non-topological
effects are averaged out, making them transparent to
the response captured by the generating functional $W[A_1, A_2]$.

Having fixed the definition of the time contour, we now discuss the structure of the gauge transformations. These have the form
\begin{align}
  \label{SW gauge trsf}
A_{1\mu}\to A_{1\mu}+\p_\mu \lambda_1,\quad
A_{2\mu}\to A_{2\mu}+\p_\mu \lambda_2\ ,
\end{align}
where $\lambda_1(t,\vec r)$ and $\lambda_2(t,\vec r)$ are independent functions, except at the end points $t_0$ and $t_1$, where they must be related as
\be \begin{split} \label{gauget}
  \lambda_1(t_0,\vec{r})&=\lambda_2(t_0,\vec{r})+2\pi n_0,
  \\
  \lambda_1(t_1,\vec r)&=\lambda_2(t_1,\vec r)+2\pi n_1 \ ,
\end{split}\ee
where $n_0$ and $n_1$ are integers. Small gauge transformations will satisfy $\lambda_1,\lambda_2\to 0$ at $t=t_0,t_1$.\footnote{
  We assume that, in the operator formalism,
  gauge transformations are implemented by
  unitary transformations of the form $V(t)=e^{i\sum_r \lambda(t,r)n_r}$, where $n_r$ is the charge density operator.
  The evolution operators in the Schwinger-Keldysh generating functional transform as
  \begin{align}
    U(t_1,t_0;A_{1}) &\to  V_1(t_1) U(t_1,t_0;A_{1})V_1^\dag(t_0),
                       \\
    \mbox{and}\quad
    U(t_1,t_0;A_{2}) &\to  V_2(t_1) U(t_1,t_0;A_{2})V_2^\dag(t_0),
  \end{align}
  which implies eq. (\ref{gauget}).}
The gauge invariance of the effective action will be further discussed in Sec.\ \ref{Response functional on the torus}.

\paragraph{Slowly-varying background}
We will restrict to background sources which are slowly varying in space and time. 
In our discussion, we will be concerned with systems which are in the
localized regime. As far as the system localizes, we expect the generating
functional $W$ to be a local functional in $A_1$ and $A_2$, which is a crucial feature of our formulation as it will allow to write down $W$ in a derivative expansion in $A_1$ and $A_2$, as far as the latter are sufficiently slowly varying, enabling us to identify particular couplings in $W$ which contribute to topological response.

\begin{center}
  $\dagger \dagger \dagger$
\end{center}

Below we will be interested in
the structure of the Schwinger-Keldysh generating functional
 $W$, which depends on background $U(1)$ gauge fields
$A_{1\mu},A_{2\mu}$
and on the constant chemical potential $\alpha$. For the rest of the paper, we will further take $A_0=0$ and
$A_i=A_i(\vec r)$, for both copies of the background. We will thus restrict to ``static'' response. Operationally, these configurations are the most general for which we can analytically compute topological response from the microscopic models that we are interested in, and take the continuum limit. We will see that this choice is sufficient to capture the topological character of periodically driven systems.\footnote{We remark that, from the point of view of our effective response, there is no technical limitation in considering terms which depend on $A_{10},A_{20}$, and which have (slow) time dependence. We leave this to future work.} Furthermore, we will focus on 2+1-dimensional systems with particle-hole symmetry, i.e. we will require
\be
\begin{gathered}
e^{iW[A_1,A_2]}=e^{iW[-A_1,-A_2]}\
\\
\nonumber
\mbox{with}\quad
\alpha \to - \alpha\ .
\end{gathered}
\ee


It is convenient to introduce a new basis for the background,
\be
A_{ri}=\frac 12(A_{1i}+A_{2i}),\quad
A_{ai}=A_{1i}-A_{2i}\ ,
\ee
where this change of basis is sometimes referred to as the Keldysh rotation
\cite{kamenev_2011}.
This basis is convenient as the constraints (\ref{cons}) can be easily
implemented.  We will write $W$ as an expansion in number of derivatives acting
on $A_{ri},A_{ai}$, and a power expansion in $A_{ai}$, which will make it
easy to enumerate the list of terms compatible with (\ref{cons}). Note that the
second condition in (\ref{cons}) requires each term in $W$ to contain at least
one power of $A_{ai}$.
To zeroth order in derivatives, there is no gauge invariant term that we can write down. To first order in derivatives, the most general generating functional is
\be
\label{theta}
W=i \frac{\Theta(\alpha)}{2\pi T}\int dt d^2 r B_a,\quad
B_a=\vep^{ij}\p_iA_{aj}\ ,
\ee
where $\Theta(\alpha)$ is an arbitrary function of $\alpha$. One immediately
sees that (\ref{theta}) satisfies conditions (\ref{cons}).
Additional terms will be at least second order in derivatives, such as
$(\vep^{ij}\p_i A_{aj})^2$ or $(\vep^{ij}\p_i A_{rj})(\vep^{kl}\p_k A_{al})$.
Since we are interested in topological responses, we will focus on (\ref{theta}),
as it is the only term with a coupling constant that is dimensionless in length units.
In the next sections we will focus on a family of systems which displays
precisely this type of response, and we will see how their topological properties are encoded
in the function $\Theta(\alpha)$.
We will look at Floquet systems defined on
closed as well as open spatial manifolds.
In the first case, we will consider backgrounds with nontrivial flux in order for (\ref{theta}) to contribute, while in the second case (\ref{theta}) can be written as a boundary term.

\section{Topological chiral Floquet drive}
\label{Chiral Floquet drive}

\begin{figure}[t]
  \begin{centering}
    \includegraphics[scale=0.6]{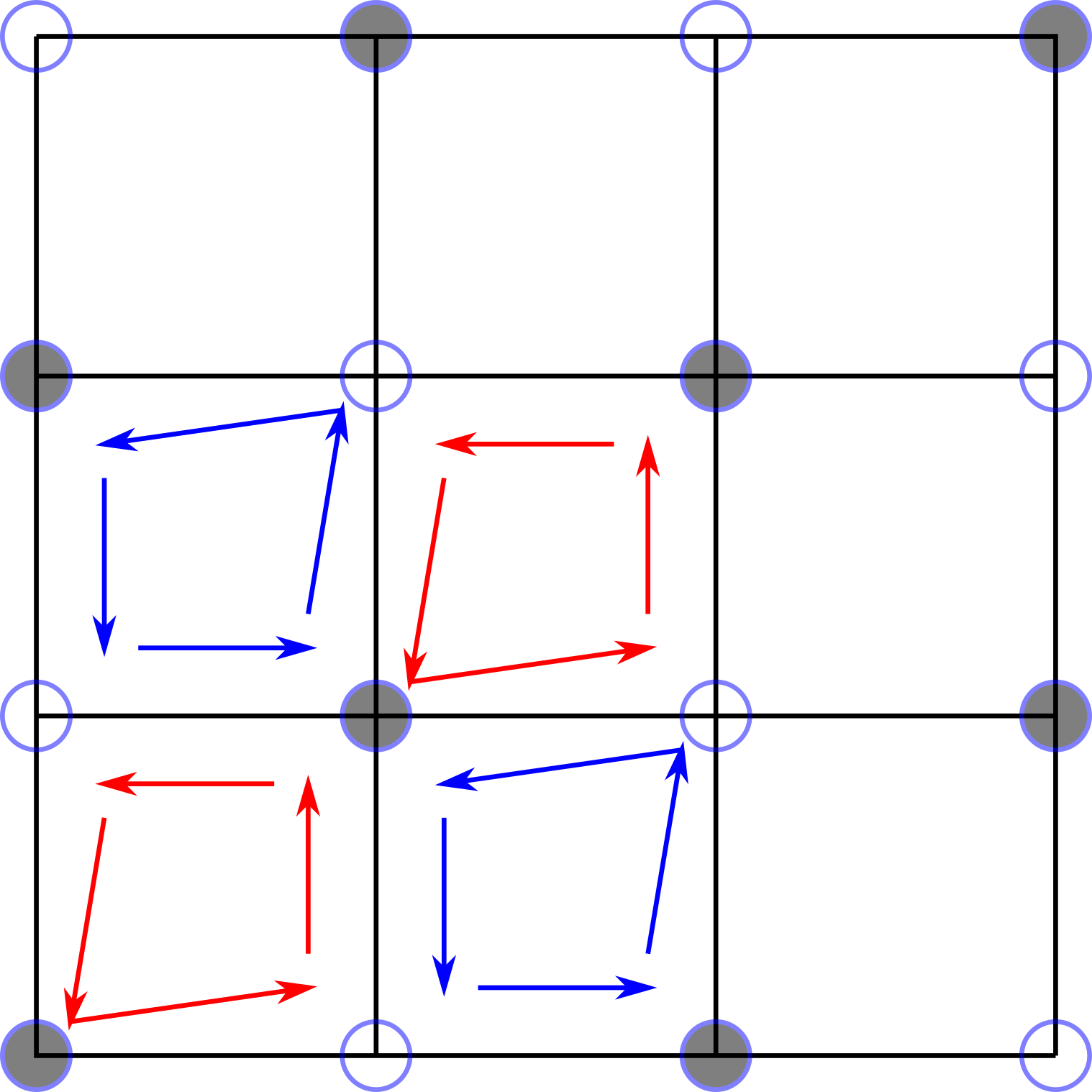}
  \end{centering}
\caption{Chiral Floquet drive.
  Red and blue arrows represent
  chiral trajectories of particles starting from
  A and B sublattice.
\label{chiralfloquet}}
\end{figure}

In this section we shall study in detail the Schwinger-Keldysh Floquet response
of a particular model.
Consider a two-dimensional square lattice with periodic boundary conditions of
size $L_x\times L_y$,
where $L_x,L_y$ are even integers.
The total number of sites is $L_x L_y =N$.
We denote site coordinates with $r=(x,y)\in \mathbb{Z}\times \mathbb{Z}$, and split
sites into sublattice $A$, with coordinates $x+y \in 2\mathbb{Z}$, and
sublattice $B$ with coordinates $x+y \in 2\mathbb{Z}+1$.
The model is given by a Floquet Hamiltonian $H(t)$ of period $T$ obtained as follows.
Divide the period $T$ in five intervals of equal duration $T/5$,
where each of the first four intervals has Hamiltonian
$H_n$, with $n=1,2,3,4$, where
\begin{align}
\begin{gathered}
  \label{hamn}
  H_n = \sum_{r\in A}H_{n,r}\ ,
  \\
  H_{n,r}=-J\left( e^{i A_{r,r+b_n}} c^{\dag}_r c^{\ }_{r+b_n} + h.c. \right)\ ,
\end{gathered}
\end{align}
with $J=\frac{2.5\pi}T$, and where
\begin{align}
  \label{bbn}
  b_1 = -b_3=(1,0),
  \quad
  b_2 = -b_4 = (0,1) \ ,
\end{align}
while during the fifth interval the Hamiltonian is zero. The fifth interval will
be of practical use later, when we shall introduce disorder. Note that the
$H_{n,r}$ and $H_{n,r'}$ commute with each other, so the evolution can be
factorized on each site $r\in A$. The resulting evolution is to move a particle
around a plaquette,
and bring it back to its original position after one period, as illustrated in
Fig.\ \ref{chiralfloquet}.
This model was originally introduced in
\cite{2012arXiv1212.3324R}
and has been extensively studied
e.g.\ in \cite{2016PhRvX...6b1013T,2017PhRvL.119r6801N}.
In addition, we added a minimal coupling to a background $U(1)$ gauge field, where
\be A_{r,r+b}=\int_r^{r+b}dr\cdot A(r)\ee
is the gauge link from site $r$ to site $r+b$.
Note that, as mentioned in Sec. \ref{Schwinger-Keldysh response}, we are restricting to background gauge fields with $A_0=0$ and $A_i=A_i(r)$. In principle, one can take slowly time-dependent background sources $A_i(t, \vec r)$ and perturbatively solve the model by doing derivative expansion in time. However, we expect such configurations to contribute only through higher derivatives to the generating functional $W$. Static configurations will be sufficient to evaluate $\Theta(\alpha)$ introduced in (\ref{theta}), thus capturing the topological character of these periodically driven systems.

The Floquet
unitary is given by
\begin{align}
  \begin{gathered}
U(T/2,-T/2) \equiv U_F = U_4 U_3 U_2 U_1,
\nonumber \\
U_n = \prod_r e^{-i T/5 H_{n,r}}
\end{gathered}
\end{align}
with
\begin{align}
  e^{-i T/5 H_{n,r}}
  &= 1-(n_r-n_{r+b_n})^2
    \nonumber \\
  &+i(n_r-n_{r+b_n})^2 \left(e^{iA_{r,n}} c^\dag_r c^{\ }_{r+b_n}+\text{h.c.}\right)
                       \ .
\end{align}
Crucially, we notice that this model has a unitary on-site particle-hole symmetry, given by
\begin{align}
  \label{PHS}
  \begin{gathered}
    c^{\ }_r \to (-1)^rc^{\dag}_r, \quad
    c^{\dag}_r \to (-1)^r c^{\ }_r,
    \nonumber \\
    A_i(r) \to -A_i(r)\ ,
  \end{gathered}
\end{align}
where $(-1)^r=+1$ or $-1$ if $r$ belongs to sublattice $A$ or $B$, respectively.
This $\mathbb Z_2$ symmetry will imply an effective theory argument for the
quantization of our response.

This Floquet drive is special or ideal in the sense that,
in the absence of the background gauge field,
$U_F=I$. I.e., the Floquet Hamiltonian is identically zero, and hence no heating.
For more generic models, this is not the case and
$U_F=\exp (-i T H_F)$,
where $H_F$ is a Floquet Hamiltonian.
To avoid heating, we need to demand $H_F$ is, e.g., many-body localizing.

\subsection{Response functional on the torus}
\label{Response functional on the torus}

We shall now compute the Schwinger-Keldysh generating functional $W[A_1,A_2]$
introduced in \eqref{genf} for the chiral Floquet drive, subject to the periodic
boundary conditions described in the beginning of this Section.
Since the Hamiltonian is quadratic ($U$ is Gaussian),
$W$ reduces to a quantity built from the
single-particle counter part of the floquet unitary;
$U(t_1,t_0)$ transforms the fermion creation/annihilation operators as
$U(t_1,t_0) c_i U^{\dag}(t_1,t_0)= \mathcal{U}_{ij}(t_1,t_0)c_j$
and
$U(t_1,t_0) c^{\dag}_i U^{\dag}(t_1,t_0) = \mathcal{U}^*_{ij}(t_1,t_0)c^{\dag}_j =
c^{\dag}_j U^{\dag}_{ji}(t_1,t_0)$
where the $N \times N$ unitary matrix $\mathcal{U}_{ij}(t_1,t_0)$
is the single-particle evolution operator acting on the single-particle Hilbert space.
By noting the formula,
\begin{align}
  \mathrm{Tr}\, \left(e^{ \sum_{i,j} c^{\dag}_i \mathcal{A}_{ij} c_j}\right)
  =
  \mathrm{det}\, \left( I + e^{ \mathcal{A} } \right),
\end{align}
we then find
\begin{align}
  \label{dett}
  &e^{iW}
  = \frac 1{\Tr\,(e^{\alpha Q}) }
  \mathrm{det}
    \left[ e^{-\frac \alpha 2}I +e^{\frac \alpha 2}\,
    \mathcal{U}(A_1)\,
    \mathcal{U}^{\dag}(A_2)
    \right]\ ,
\end{align}
where we used \eqref{rho} as initial density matrix,
and $\mathcal{U}(A)\equiv \mathcal{U}(\kappa T,-\kappa T;A)$.
(Here and henceforth, ``$\tr$'' and ``$\det$''
denote the trace and determinant
in the $N$-dimensional single-particle Hilbert space, respectively,
as opposed to ``$\Tr$'' which is the trace taken over
the $2^N$-dimensional many-body Hilbert space.)
We used a ``particle-hole symmetrized'' definition of number operator, i.e.
\begin{align}
  Q= \sum_r (n_r-1/2),\quad n_r=c^\dag_r c^{\ }_r\ ,
\end{align}
where the eigenvalues of $Q$ run from $-\frac N2$ to $\frac N2$, and
$\Tr\, (e^{\alpha Q})=\prod_r
(e^{-\frac \alpha 2}+e^{\frac \alpha 2})
=(2\cosh\frac \alpha 2)^{N}$.
The chemical potential $\alpha$ can be used to project the ``unnormalized''
generating functional
to a given sector with fixed particle number:
\begin{align}
  \Tr\, (e^{\alpha Q})\, e^{iW}=\sum_q e^{\alpha q} Z_{q+N/2}[A_1,A_2]\ ,
\end{align}
where the subscript $q+N/2$ is the non-symmetrized particle number running from $0$ to $N$. For the case of Gaussian Floquet unitaries, expanding the determinant in \eqref{dett} we obtain, for example,
\begin{align}
  \label{part func for q}
  Z_{1}[A_1,A_2]&=\tr\, [\, \mathcal U (A_1)\, \mathcal U^\dag (A_2)]\ ,
                \nonumber \\
  Z_{N}[A_1,A_2]&=\det\, [\,\mathcal U (A_1)\, \mathcal U^\dag (A_2)]\ .
\end{align}

It is easy to check that $\mathcal{U}_F(A)$ is diagonal with its diagonal
elements given by $e^{i B_{r}}$ where $B_r$ is a flux picked up by a particle
which is located initially at $r$:
$\mathcal{U}_F(A) = \sum_r e^{ i B_r}|r\rangle \langle r|. $
Then,
\begin{align}
  e^{iW}
  &= \frac{1}{(2\cosh\frac \alpha 2)^{N}}
    \nonumber \\
  &
    \quad \times
    \sum_{\{n_r = 0,1\}}
    e^{\left(n_r-\frac{1}{2}\right)\alpha}e^{ +i\int \frac{dt}{T} \sum_r (B_{1r}-B_{2r}) n_r}
    \nonumber \\
  &=\frac{1}{(2\cosh\frac \alpha 2)^{N}}
    \prod_r
    \left[
      e^{-\frac \alpha 2}+e^{\frac \alpha 2}e^{ i \int \frac{dt}{T}(B_{1r}-B_{2r})}
    \right]\ .
    \label{res1}
\end{align}
Note that the time integral in the exponents $\int dt$ should be thought of as
$\int_{-{\kappa}T}^{{\kappa}T}dt$, with ${\kappa}$ a sufficiently large integer.
We can check that \eqref{res1}
is consistent with particle-hole symmetry,
\begin{align}
  \label{PH bulk}
  \frac{Z[-A_1, -A_2, -\alpha]}{Z[A_1, A_2, \alpha]}
  =
  e^{ i \int \frac{dt}{T}\sum_r [B_{1r} - B_{2r}]}
  =1,
\end{align}
where we noted the quanitzation
of the total flux $\sum_r B_{sr} =2\pi \times ({\rm integer})$
$(s=1,2)$
on a
close manifold since $A_{r,r'}$ is an angular variable,
$A_{r,r'}\equiv A_{r,r'}+2\pi$.

Equation \eqref{res1} is the exact microscopic result
and can be used to study systems with arbitrary configurations
of the background gauge fields --
See around Eqs.\ \eqref{monopole response} and \eqref{gauge unit flux},
for example.
We now specialize to background configurations which are slowly varying compared
to the lattice constant.
In this limit, $\p A_i/\p r_j \ll A_i$, i.e. one expands $e^{iB_r}=1+i B_r+\cdots$, and resumming, the only finite contribution to the generating functional will be\footnote{Note that the continuum limit should be taken \emph{before} the infinite time limit, i.e. in taking $B_r\to 0$, the integral $\int dt$ should be performed over a finite time interval.}
\begin{align}
  &
  \exp{iW[A_1, A_2]}
  \nonumber \\
  &
  =
    \exp {i \frac{\Theta(\alpha)}{2\pi}
    \int \frac{dt}{T}\int d^2 r
    [B_1(r)-B_2(r)]}\ ,
\label{SK top action chiral floquet}
\end{align}
with
\begin{align}
  \begin{gathered}
  \Theta(\alpha) = \theta + f(\alpha),
  \nonumber \\
  \theta = \Theta(0) = \pi,
  \quad
  f(\alpha)=- \pi \tanh \frac \alpha 2\ .
\end{gathered}
\end{align}
Note that the generating functional is now a pure phase,
and topological in the sense that it does not require
(spatial) metric for its definition.

The effective action \eqref{SK top action chiral floquet}
is a Schwinger-Keldysh analogue of the
theta term,
$\exp[ i \frac{\theta}{2 \pi} \int_{M_2} dA]$,
which appears, e.g. as an effective
response functional of (1+1)-dimensional
static topological insulators (e.g., the SSH model),
where $M_2$ is the (1+1)-dimensional spacetime
\cite{2008PhRvB..78s5424Q}.
For the static case,
$\theta$ is a periodic variable,
$\theta \equiv \theta + 2\pi$,
because of the Dirac quantization condition:
for any (1+1)-dimensional closed Euclidean spacetime $M_2$,
$\int_{M_2} dA = 2\pi \times {\rm integer}$,
which is a consequence of the large $U(1)$ gauge invariance.
Imposing a discrete particle-hole symmetry
quantizes $\theta$ to be $\theta={\rm integer}\times \pi$,
and this then ``predicts''
symmetry-protected topological phases
(phases which are not smoothly connected to each other)
protected by particle-hole symmetry.
In other words,
the theta term, once quantized by symmetry, serves as
a topological invariant which can be used to
distinguish/label topologically distinct particle-hole symmetric
phases.

For the Schwinger-Keldysh functional
\eqref{SK top action chiral floquet},
the situation seems more complicated in the sense that
the combination $A_{ai}=A_{1i}-A_{i2}$
entering in \eqref{SK top action chiral floquet} is ``neutral'' under spatial large gauge transformations, i.e. transformations which allow a nontrivial flux of $A_{ai}$ across the torus. Indeed, due to (\ref{gauget}), $\lambda_1(t_0,r)$ and $\lambda_2(t_0,r)$ must be topologically equivalent as spatial functions, as well as $\lambda_1(t_1,r)$ and $\lambda_2(t_1,r)$,
and hence there is no large gauge transformation to
quantize $\int dA_a=\int (dA_1 -dA_2)$
(where we consider the integral only over the space,
as $\int dt/T$ is simply an integer).
Nevertheless,
we can still argue that $\Theta(\alpha)$
in \eqref{SK top action chiral floquet}
has a periodicity
$\Theta(\alpha) \equiv \Theta(\alpha)+2\pi$.
First we note that, if the dependence on $A_1$ and $A_2$
of the generating functional $Z[A_1,A_2]$
enters through the total fluxes $\sum_r B_{1r}$ and $\sum_r B_{2r}$,
they have to be separately quantized since $A_1$ and $A_2$ are
angular variables, $A_{sr}\equiv A_{sr}+2\pi$.
Second, periodicity of $\Theta(\alpha)$ can also be proven from the following argument. When we switch off
one of the gauge fields, $A_2$, say,
the Floquet unitary of the model reduces to identity
$U(\kappa T,-\kappa T;A_2=0)=I$,
i.e., the second Schwinger-Keldysh copy simply disappears.
Hence the Schwinger-Keldysh trace \eqref{genf}
reduces to
$
e^{iW[A_1, A_2=0]} = \mathrm{Tr} \left[U(A_1) \rho_0 \right]
$
which is now invariant under the ``accidental'' large gauge transformation
$A_{1i} \to A_{1i} + \partial_i \lambda$, while $A_{2i}$ remains zero.\footnote{This enhanced symmetry can also be seen without setting $A_{2}=0$ by simply noting that $U(A_2)$, for this particular model, commutes with gauge transformations.}
On the other hand,
the effective action \eqref{SK top action chiral floquet} reduces to
$
\exp [
-i \frac{\Theta(\alpha)}{2\pi}\int \frac{dt}{T}\int d^2 r B_1(r)
]
$
in which $\Theta(\alpha)$ should be now periodic
because of the (large) gauge invariance under
$A_{1i} \to A_{1i} + \partial_i \lambda$.

In the latter argument above, we relied on a special feature of the model,
$U( {\kappa}T,\kappa T;A=0)=I$,
which is not true in general:
for more general cases
$U( {\kappa}T,-\kappa T;A=0)$ is not the identity,
by given by
the exponentiated Floquet Hamiltonian,
$U({\kappa}T,-\kappa T;A=0) = \exp (-i {2\kappa}T H_F)$.
Nevertheless, the periodicity of the theta angle $\Theta(\alpha)$
in the Schwinger-Keldysh effective action will persist at least for a wide class of models,
thanks to the fact that the value of $\Theta(\alpha)$ is independent of
continuous deformations of the system,
as will be shown in Sec.\ \ref{sec:stabi}.
 Indeed, Floquet unitaries $U(t)$ can be smoothly deformed into the form
$U'(t)= \tilde{U}(t)\exp (-i t H_F)$,
where $\tilde{U}(t)$ is periodic (the so-called micro motion part),
 with $\tilde{U}(t=T)=I$, and $\exp (-i t H_F)$ captures the non-periodic part.
 If one can then smoothly deform $U(t)$ into $\tilde U(t)$ (see e.g.\
 \cite{2017PhRvB..95s5128R}),
which one can do with the non-periodic evolutions of the models discussed in
later sections\footnote{In fact, some literature just entirely focuses on
periodic Floquet unitaries $U(0)=U(T)=I$,
see Ref.\ \cite{2017PhRvB..95s5128R}.},
the identification $\Theta(\alpha)=\Theta(\alpha)+2 \pi$ continues to hold.


Now, under particle-hole symmetry \eqref{PHS},
$(A_{1i},A_{2i},\alpha)\to (-A_{1i},-A_{2i},-\alpha)$, i.e.,
\begin{align}\label{phcon}
& \exp{iW[A_1,A_2]}
  \nonumber \\
  & \to \exp{
-i \frac{\Theta(-\alpha)}{2\pi}\int \frac{dt}{T}\int d^2 r [B_1(r)-B_2(r)]}\ ,
\end{align}
which means that, when $\alpha=0$,
$\theta$ is quantized as $\theta=\pi \times {\rm integer}$.
As in the case of static topological insulators,
the quantized theta term can be thought of as a topological
invariant differentiating topologically distinct
(many-body localized) Floquet unitaries
(regardless of the microscopic details of the system,
and even for strongly coupled many-body systems, as far as the thermodynamic
limit is well-defined).
For generic values $\alpha$, one can see that particle-hole symmetry implies
$f(\alpha)=-f(-\alpha)$.
In the next Section, we will show that $\Theta(\alpha)$ is independent of
continuous deformations of the Hamiltonian,
and that $f(\alpha)$ contains additional topological information of the system.

We close this subsection with a few remarks.
First, while we have been focusing on smooth configurations of the background
gauge fields, it is also interesting to consider non-smooth configurations,
e.g.,
a pair of localized magnetic fluxes $\phi$ and
$-\phi$ inserted through two plaquettes.
The corresponding background gauge field can be introduced by
considering a ``string'' on the dual lattice connecting these two plaquettes,
and assigning $e^{i A_{rr'}}=e^{\pm i \phi}$
for those links intersecting the string.
It is straightforward to see
\begin{align}
  \label{monopole response}
Z[A_1, A_2=0] = (1/2)(1+\cos \phi)
\end{align}
where we set $\alpha=0$ for simplicity.
The partition function is real and its amplitude is zero for $\phi =
\pi \times {\rm integer}$.
This background configuration is fairly singular,
and cannot be described by the topological effective action.
The situation is similar to the response effective action of
the (integer) quantum Hall effect;
in the presence of the Chern-Simons term, the response partition function
vanishes when one introduces a monopole.
(See \cite{PhysRevLett.66.276} for example.)

Second, while we have been discussing the free fermion model,
the topological response functional \eqref{res1}
can be also derived for more generic models.
Consider the floquet models introduced and discussed in Refs.\
\cite{2016PhRvX...6d1070P,2017PhRvL.118k5301H}.
These models consist of swap operators, acting on each link.
As an example, we follow \cite{2017PhRvL.118k5301H}.
The model consists of hard-core bosons living on a square lattice.
For each link, we define a SWAP operator,
\begin{align}
 S_{r, r'} |n_r, n_{r'}\rangle = |n_{r'}, n_r\rangle
\end{align}
where $n_r = 0,1$ is the occupation number of hard core bosons
at site $r$. $S_{r, r'}$ can be given explicitly as
\begin{align}
  S_{r, r'} = 1 + b^{\dag}_r b^{\ }_{r'} + b^{\dag}_{r'} b^{\ }_r
  - n_r - n_{r'} + 2 n_r n_{r'}.
\end{align}
Combing these SWAP operators,
$
  U_j
  =
  \prod_{{r} \in A} S_{{r}, {r}+{b}_j}
$,
the total Floquet drive is given by
$
U_F= U_4 U_3 U_2 U_1
$.
In the absence of boundaries, one can readily check that
$U$ is the identity operator,
\begin{align}
 \langle \{n\} | U_F| \{n'\}\rangle
  =
  \delta_{\{n\}, \{n'\}}
  =
  \prod_{r} \delta_{n_r, n'_r}.
\end{align}
The background $U(1)$ gauge field can be introduced
by replacing
$b^{\dag}_r b^{\ }_{r'}\to b^{\dag}_r g_{rr'}b^{\ }_{r'}$
in $S_{rr'}$
where $g_{rr'}=g^*_{r'r}$ and $g_{rr'}= e^{i A_{rr'}}\in U(1)$.
One can check easily
$
  S_{r r'}(A)|n_r, n_{r'}\rangle
  =
  g^{n_r - n_{r'}}_{rr'} |n_{r'}, n_r\rangle.
$
$U_F(A)$ is diagonal
in the occupation number basis and given by;
\begin{align}
 \langle \{n\} | U_F(A)| \{n'\}\rangle
  =
  \exp [i I(n, A)]
  \delta_{\{n\}, \{n'\}}.
\end{align}
Here, for a fixed configuration $\{n\}$,
$e^{ i I (n, A)}$ can be written as
\begin{align}
  e^{ i I (n,A)}
  =
  \prod^{n_r=1}_{r} e^{ i B_r}
  =
  \prod_{r} e^{ i B_r n_r}
  =
  e^{i \sum_r B_r n_r}
\end{align}
where the product $\prod_r^{n_r=1}$ is over all $r$ where
a particle is present, $n_r=1$.
It is then straightforward to see that
the topological response functional
is given by \eqref{res1}.

\begin{figure}[t]
  \begin{centering}
    (a)\hphantom{AAAAAAAAAAAAAAAAAAAAAAAA}\\
    \includegraphics[scale=0.2]{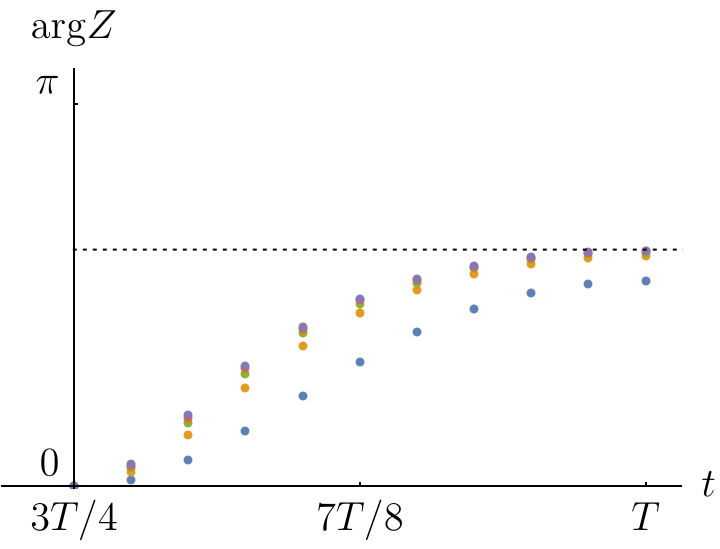} \\
    (b)\hphantom{AAAAAAAAAAAAAAAAAAAAAAAA}\\
    \includegraphics[scale=0.2]{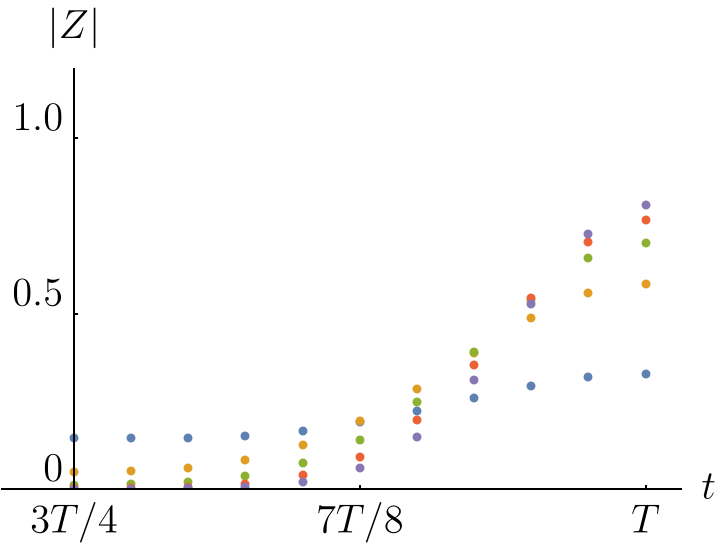}
  \end{centering}
\caption{
  The time-evolution of $Z[A_1,A_2]$ for
  the static gauge field configuration
  \eqref{gauge unit flux}
  for $3T/4 < t < T$
  for $L_x=L_y=4,8,12,16,20$
  with $\alpha=0.8$ (from the bottom at $t=T$).
  The dotline represents
  $\Theta(\alpha=0.8)=1.94795$.
  \label{numerics}
}
\end{figure}

Third, while we have focused for the Floquet unitary at
$t_1-t_0={\rm (integer)}\times T$, we can
monitor the time-evolution of the partition function
$Z[t_1,t_0;A_1,A_2]$ numerically
for arbitrary $t_{0,1}$ and
for a given static gauge field configuration.
In Fig.\ \ref{numerics}, $Z[t_1,t_0=-T/2;A_1,A_2]$
is plotted as a function of $t_1$
for the background field configuration $A_{2i}=0$ and
\begin{align}
  \label{gauge unit flux}
  &A_{1x} (x,y) = 0,
    \nonumber \\
  &A_{1y} (x,y) =
    \left\{
    \begin{array}{ll}
      \displaystyle
      0, & y = 1, \ldots, L_y -1, \\
      \displaystyle
      \frac{2\pi x}{L_x}, & y = L_y.
    \end{array}
                            \right.
\end{align}
In this configuration, the magnetic flux
is inserted through plaquette located on a row at $y=L_y$. The total flux is $2\pi$.
We see that the amplitude $|Z[A_1,A_2]|$ approaches to $\sim 1$
as $t_1\to T/2$.
On the other hand, away from $t_1=T/2$, the amplitude $|Z[A_1,A_2]|$ can be very
small (nearly zero);
in these time regions, $Z[A_1,A_2]$ seems not to be topological in nature.
In addition, as $t_1 \to T/2$, $\mathrm{arg}\, Z\to \Theta(\alpha)$.

\subsection{Open boundary conditions and magnetization}

\begin{figure}[t]
  \begin{centering}
    (a)\hphantom{AAAAAAAAAAAAAAAAAAAAAAAA}\\
    \includegraphics[scale=0.5]{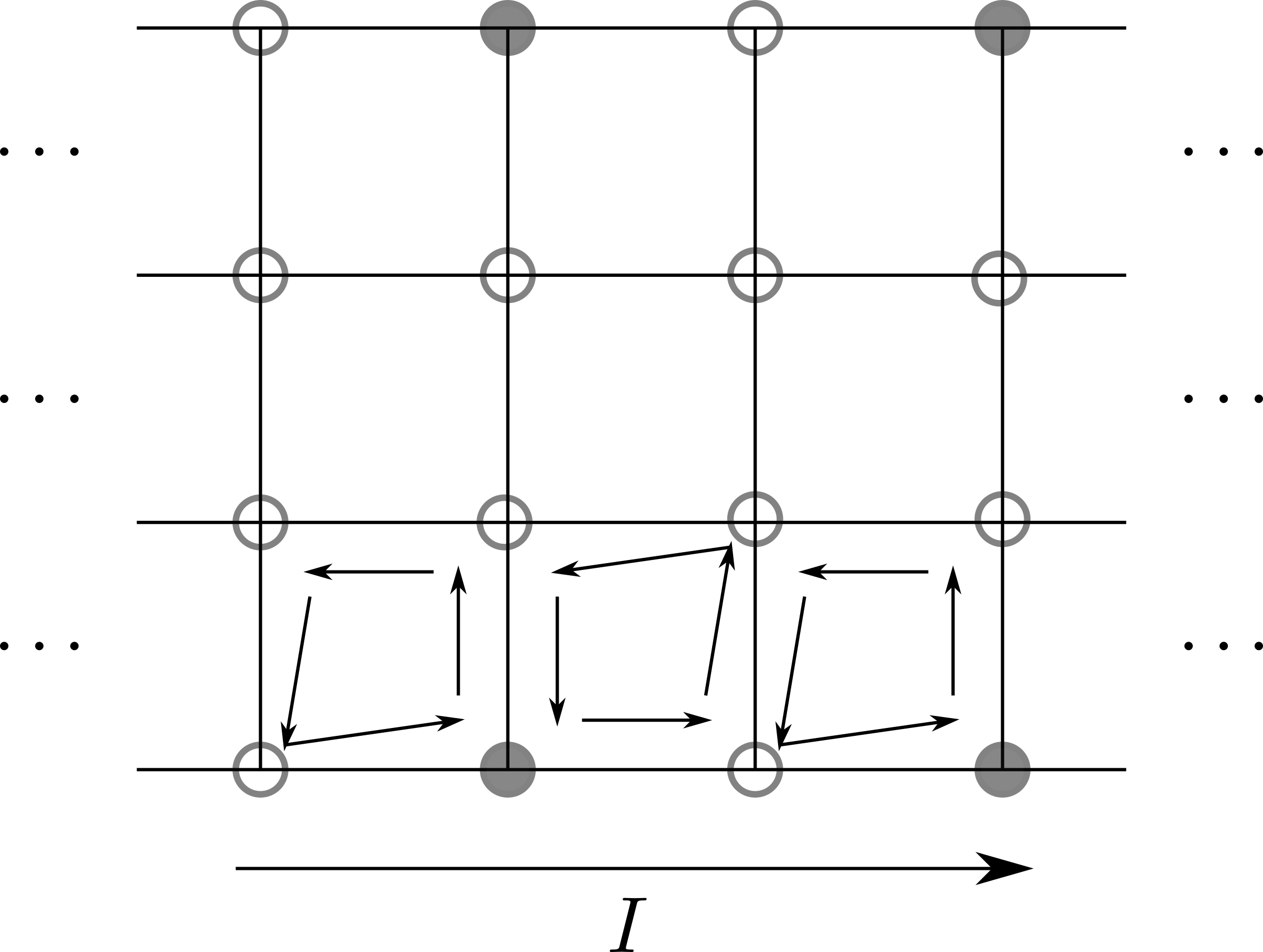}\\
    (b)\hphantom{AAAAAAAAAAAAAAAAAAAAAAAA} \\
    \includegraphics[scale=0.5]{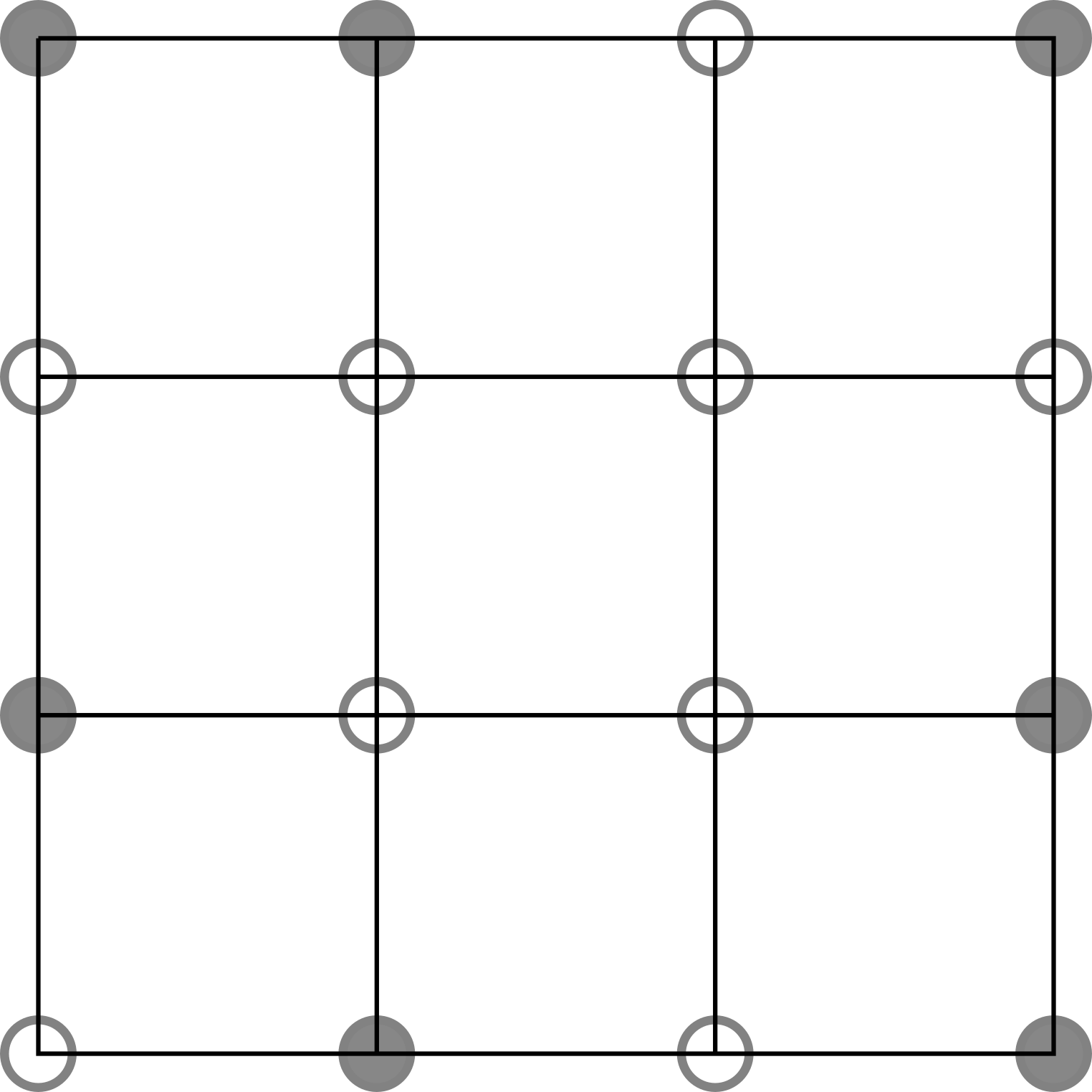}
  \end{centering}
\caption{Chiral Floquet drive with open boundary.
  (a) Cylinderical geometry with open (periodic) boundary condition in $y$ ($x$)
  direction,
  and
  (b) ``Disc'' geometry with open boundary condition in both $x$ and $y$ directions.
  Shaded (unshaded) sites belong to the boundary (bulk) Hilbert space.
\label{chiralfloquet2}
}
\end{figure}

\subsubsection{Separating bulk and boundary unitaries}
We now move on to discuss the chiral Floquet drive in the presence of open
boundary conditions.
Let us first recall that, in the absence of boundaries,
and with background gauge field, the single-particle unitary
$\mathcal{U}_F(A)$ is diagonal in the occupation number basis,
with the diagonal elements depending on the background $A$.
Let us now make a boundary by removing some links.
While the bulk part of $\mathcal U_F(A)$ continues to be diagonal, the boundary part is not, as after one period
the location of a particle on the boundary is shifted,
see Fig.\ \ref{chiralfloquet2}. We can then decompose $\mathcal{U}_F(A)$ as
\begin{align}
  \label{decompose bulk and bdry}
  \mathcal{U}_F(A)
  = \mathcal{U}_{\text{bulk}}(A)
  \oplus \mathcal{U}_{\text{bdry}}(A),
\end{align}
where
$\mathcal{U}_{\text{bulk}}$ and $\mathcal{U}_{\text{bdry}}$ are supported
by two spaces orthogonal to each other;
we will refer them as the bulk and boundary Hilbert spaces.
For our current model, the boundary
Hilbert space consists of a subset of sites living on the boundary,
as in Fig.\ \ref{chiralfloquet2}.
Correspondingly, the many-body Floquet unitary factorizes,
$U_F=U_{\text{bulk}}\otimes U_{\text{bdry}}$,
leading to the factorization of the generating functional:
\begin{align}
  \label{gen3}
  e^{iW[A_1, A_2]}
  &=\Tr\,[U(A_1)\rho_0 U^\dag(A_2)]
  \nonumber \\
  & =\Tr\,[U_{\text{bulk}}(A_1)\rho_{0,\text{bulk}} U_{\text{bulk}}^\dag(A_2)]
  \nonumber \\
  & \quad
    \times
    \Tr\, [U_{\text{bdry}}(A_1)\rho_{0,\text{bdry}} U_{\text{bdry}}^\dag(A_2)]
    \nonumber \\
  &= e^{iW_{\text{bulk}}[A_1,A_2]}
e^{iW_{\text{bdry}}[A_1,A_2]}\ ,
\end{align}
where we also split the initial density matrix into bulk and boundary parts:
$\rho_0=\rho_{0,\text{bulk}}\otimes \rho_{0,\text{bdry}}$.

The bulk effective functional $W_{\rm bulk}[A_1,A_2]$ can be
computed in the same way as the torus case,
and is essentially given by \eqref{res1},
where now in the product $\prod_r$ we simply remove sites which belong
to the boundary Hilbert space.
Taking the continuum limit,
\begin{align}\label{wbulk}
  W_{\text{bulk}}[A_1,A_2]
  &=
    \frac{\Theta(\alpha)}{2\pi}\int \frac {dt}{T}\int_{\text{bulk}} d^2 r\,
    [B_1(r)-B_2(r)]\ .
\end{align}
We also note that
\begin{align}
  \frac{Z_{\rm bulk}[-A_1, -A_2, -\alpha]}{Z_{\rm bulk}[A_1, A_2, \alpha]}
  =
  e^{i \int \frac{dt}{T}\sum^{\rm bulk}_r (B_{1r}-B_{2r}) }
  \neq 1,
\end{align}
since the total flux $\sum^{\rm bulk}_r B_r$ on an open manifold
is not subject to the quantization condition;
particle-hole symmetry is broken.

The effective functional for the boundary unitary $W_{{\rm bdry}}$ can also be
evaluated directly.
Note however that in contrast to the bulk unitary
the boundary unitary is not gapped (many-body localized),
$U_{{\rm bdry}}(A=0)$ is not identity,
and hence we do not expect $W_{{\rm bdry}}$ to be local;
we do not write down $W_{\text{bdry}}$ explicitly here.\footnote{
  The Schwinger-Keldysh generating functional in the $N$-particle sector
  takes a simple form and is given by
  \begin{align}
    Z_{N, {\text{bdry}}}[A_1,A_2]=
    e^{ -i\int\frac{dt}{T} \modtwosum_{{\rm bdry}} (A_1-A_2)}.
  \end{align}
}
Nevertheless, one can verify
\begin{align}
  \frac{Z_{\rm bdry}[-A_1, -A_2, -\alpha]}{Z_{\rm bdry}[A_1, A_2, \alpha]}
  =
  e^{ -i \int\frac{dt}{T}\modtwosum_{{\rm bdry}} (A_1-A_2) }
  \neq 1
\end{align}
where $\modtwosum_{{\rm bdry}}$ represents the sum
taken over links on the boundary region
(an analogue of a 1d line integral $\oint$ along the boundary),
and $L$ is
the circumference of the boundary.
The ``Wilson loop''
$e^{ -i\int\frac{dt}{T} \modtwosum_{{\rm bdry}} (A_1-A_2) }$
is not subject to quantization condition, and hence
particle-hole symmetry is broken, as in the bulk.
On the hand, when the bulk and boundary effective functionals
are combined, the total effective functional respects the
particle-hole symmetry,
$
{Z[-A_1, -A_2, -\alpha]}/{Z[A_1, A_2, \alpha]} =1
$.
The situation is similar
for the trace of the single unitary operator
$\mathrm{Tr}\, [U_{\rm bdry}(A)]$ (which is not the
Schwinger-Keldysh trace),
which takes a simple form and is given by
\begin{align}
  \mathrm{Tr}\, [U_{\text{bdry}}(A)]
  = e^{- L \frac{\alpha}{2}}
  + e^{+L \frac{\alpha}{2}}
  e^{ i \modtwosum_{\text{bdry}} A}
  \label{tr single}
\end{align}
where $L$ is
the total number of sites in the boundary Hilbert space.
The trace \eqref{tr single} does not
preserve particle-hole symmetry, while it enjoys the large $U(1)$ gauge
invariance.
On the other hand, by adding (multiplying) a counter term,
$
e^{ -(i/2)\int\frac{dt}{T} \modtwosum_{\text{bdry}} A}\mathrm{Tr}\,
[U_{\text{bdry}}(A)]
$
is particle-hole symmetric, but not
invariant under large $U(1)$ gauge transformations.
The situation is completely analogous
to the well-known mixed anomaly
(a conflict between particle-hole and $U(1)$ symmetry)
in (0+1)-dimensional field theory
\cite{Elitzur:1985xj}.
This is consistent with the fact that the boundary unitary
realizes a single chiral (Weyl) fermion;
the single particle boundary unitary in momentum space is given
simply by $\mathcal{U}_{{\rm bdry}}(A) = \exp i k_x$ where
$k_x$ is single particle momentum along the boundary.

Two comments are in order.
First, the bulk response has the same form as that of the closed system
discussed in the previous subsection.
This is expected as the system is localized.
In other words, thanks to localization, particle-hole invariance implies that
$\theta$ is quantized even for the open system.
We will support this statement with more general models
in Sec.\ \ref{Stability under disorder and perturbations}.
Second, the value of $\theta$ is unambiguously defined,
while in the case of periodic boundary conditions it is defined only mod $2\pi$.
This has a well-known counter part in the context of static SPT phases, such as
topological insulators.

\subsubsection{Magnetization}
Since the magnetic flux can have a continuous value,
we can differentiate $W_{\text{bulk}}[B]$ with respect to $B$ and directly relate our response to magnetization. Indeed,
\begin{align}
  \label{magn}
  &
  \left.\frac{\partial}{\partial B}e^{iW_{\text{bulk}}[B]}\right|_{B=0}
    \nonumber \\
  \quad
  &=
    \mathrm{Tr}\,
    \left[\rho_0
    U^{\dag}(B=0) \frac{\partial}{\partial B} U (B)
    \right]
    \nonumber \\
  &=
    -i \int^{{\kappa}T}_{-{\kappa}T} dt\,
    \mathrm{Tr}\,
    \left[\rho_0
    U^{\dag}(t,-{\kappa}T)
    \frac{\partial H(t)}{\partial B} U(t,-{\kappa}T)
    \right]
    \nonumber \\
  &=    -i \int^{ {\kappa}T}_{-{\kappa}T} dt\,
    \mathrm{Tr}\,
    \left[\rho_0
    U^\dag(t,-{\kappa}T)M U(t,-{\kappa}T)
    \right]
\end{align}
where we suppressed the subscript bulk from various quantities for simplicity,
${\kappa}$ is a half-integer, and we used
\begin{align}
  &
    \left.U^{\dag}( {\kappa}T,-{\kappa}T) \frac{\partial}{\partial B} U ({\kappa}T,-{\kappa}T;B)\right|_{B=0}
    \nonumber \\
  &\quad
    =
    -i \int^{{\kappa}T}_{-{\kappa}T} dt\, U^{\dag}(t,-{\kappa}T)
    \frac{\partial H(t,B)}{\partial B} U(t,-{\kappa}T),
\end{align}
and where we identify $M \equiv -\partial H/\partial B$ as the magnetization
operator.
We are then led to introduce
\begin{align}
  m_\alpha=
  \frac{i}{2 {\kappa}T L_x L_y}\left.\frac{\partial}{\partial B}e^{iW_{\text{bulk}}[B]}\right|_{B=0}\ ,
\end{align}
where the factor of $2 {\kappa}T$ is the total length of the time integral,
and $L_x L_y$ is the area of the bulk.
We naturally view $m_\alpha$ as the magnetization averaged over time and space.
This quantity was introduced in
\cite{2017PhRvL.119r6801N}
for the single particle ``infinite temperature'' state.
Using the bulk generating functional worked out in (\ref{wbulk}), we then find
\begin{align}
  m_\alpha= -\frac{\Theta (\alpha)}{2\pi T}\ ,
\end{align}
so that, for $\alpha=0$, the averaged magnetization is half-quantized.
In \cite{2017PhRvL.119r6801N}
it was found that the averaged magnetization is quantized.
The relative factor of $1/2$ is in that we are considering the sum over states
with arbitrary particle numbers.

If we focus on the $N$-particle sector of the Hilbert space,
we have the integral quantization of the averaged magnetization.
Indeed, explicit evaluation of the generating functional restricted to the $N$-particle sector $Z_N[A_1,A_2]$ in \eqref{part func for q} gives
\begin{align}
    \label{bulk resp}
  &- i \log Z_{N,\mathrm{bulk}}[A_1, A_2]
    \nonumber \\
  &\quad
    =
     \frac{2 \theta}{2\pi}
    \int \frac{dt}{T}\int d^2r\, [B_1(r)-B_2(r)],
\end{align}
for smooth background gauge fields. Note the relative factor of 2
as compared to \eqref{wbulk}. Using again eq. (\ref{magn}), where this time $\rho_0$ is the density matrix supported on the bulk and restricted to particle number $N$, gives
\be \frac{\p}{\p B} Z_{N,\text{bulk}}[A,0]=-i\int_{-\kappa T}^{\kappa T}dt\,{\Tr}'  M(t) \ ,\ee
where  ${\Tr}'$ is $N$-particle trace taken over the bulk sites. This then gives
the time-averaged magnetization per unit area as
\begin{align}
  \frac{1}{2 \kappa L_x L_y}
  \int^{\kappa T}_{-\kappa T} dt\, {\Tr}'  M(t)
  =
   \frac{\theta}{\pi}\ .
\end{align}

\subsection{Stability under deformations}\label{sec:stabi}

We shall now show that $\Theta(\alpha)$ must be independent of continuous
deformations of the Hamiltonian, as far as the system is localized.
In any geometry, such as the torus described in
Sec.\ \ref{Response functional on the torus}
or the strip of Fig.\ \ref{strip},
consider smoothly deforming the Hamiltonian inside two regions $I$ and $II$
whose size and distance
is much larger than the localization length, and denote by $H_I(t),H_{II}(t)$
the Hamiltonian in region $I,II$, respectively.
Further, assume that the length scale of deformation from $H_I(t)$ to
$H_{II}(t)$
is much shorter than the scale of
variation of the gauge field $A_i$.
The response at first derivative order must then be
\begin{align}
  \label{gamma}
  W[A_a]=\frac{1}{2\pi }\int \frac{dt}{T} \int d^2 r\, \Theta(\alpha,r)\, \vep^{ij}\p_i A_{aj}(r)\ ,
\end{align}
where $\Theta(\alpha,r)$ approaches the value
$\Theta_I(\alpha),\Theta_{II}(\alpha)$ in region $I,II$, respectively, and the chemical potential $\alpha$ is constant everywhere.
Varying the generating functional with respect to $A_i(r)$ gives the time-averaged expectation value of the current,
\begin{align}
  &\left.\frac{\delta e^{iW[A]}}{\delta A_i(r)} \right|_{A=0}
  \nonumber \\
  &=-i\int dt \Tr\left[\rho_0 U^\dag(t,-{\kappa}T)\frac{\p H(t)}{\p A_i(r)}U(t,-{\kappa}T)\right]
  \nonumber \\
  &=-i\int dt\Tr[\rho_0 J^i(r,t)]\equiv-i\bar J^i(r)\ ,
\end{align}
where we used steps similar to those around Eq.\ \eqref{magn}.
Plugging in the functional (\ref{gamma}) gives
\begin{align}
  \bar J^i(r)=-\frac 1{2\pi }\int \frac{dt}{T}\,
  \vep^{ij}\p_j\Theta(r)\ .
\end{align}
Due to localization this current should vanish, as far as $r$ is sufficiently
far from any boundaries,
such as the boundary of the strip in Fig.\ \ref{strip}, or the boundary of the
cylinder itself.
We now show why this is the case for a model of the form
$H(t)=H_0(t,A)+H_{\text{int}}(t)$,
where $H_0$ is the chiral Floquet Hamiltonian in (\ref{hamn})-(\ref{bbn}),
and $H_{\text{int}}(t)$ is a generic interaction term which does not depend on
$A_i$, and has a generic time dependence, i.e. it does not have to be periodic: any $H_{\text{int}}(t)$ will be fine as far as the system remains many-body localized. The trace of the current operator for the Hamiltonian $H(t)$ evaluated at $r=\bar r$ is
\begin{align}
  -i\Tr[\rho_0 J^i(\bar r,t)]=\Tr[\rho_0U^\dag(t)(c^\dag_{\bar r+i}c_{\bar r}-c_{\bar r}^\dag c_{\bar r+i})U(t)]\ .
\end{align}
If the system is on a closed manifold, where $\rho_0$ does not project out any
states, $\rho_0$ commutes with $U(t)$ which immediately leads to the vanishing
of the trace,
$\Tr\, [\rho_0U^\dag(t)(c^\dag_{\bar r+i}c^{\ }_{\bar r}-c_{\bar r}^\dag c^{\ }_{\bar r+i})U(t)]
=\Tr\, [\rho_0(c^\dag_{\bar r+i}c^{\ }_{\bar r}-c^{\dag}_{\bar r} c^{\ }_{\bar r+i})]
=0
$.
(See \cite{2019arXiv190712228N} for a similar discussion.)
Thus, as far as $W[A]$ is given by the local functional
\eqref{gamma}, $\Theta(\alpha)$ must be independent of
continuous deformations.
In the presence of boundaries,
such as the geometry similar to that of Fig.\ \ref{strip},
one can still factorize the total unitary
into its bulk and boundary parts
(c.f. \eqref{decompose bulk and bdry}).
While the boundary unitary is not many-body localized,
as far as the current operator is evaluated at location $\bar r$
well inside the bulk region, we conclude that
the trace of the current operator should still be zero,
which then implies that $\Theta$ cannot be changed continuously. Note that in the above proof we did not make any use of the periodicity of the Hamiltonian. Independence on continuous deformations of this topological response is guaranteed solely by localization.

\subsection{Numerical tests of stability}
\label{Stability under disorder and perturbations}

As mentioned in the beginning of this section,
the topological chiral Floquet model
\eqref{hamn}-\eqref{bbn}
is somewhat special or ideal in the sense that its Floquet Hamiltonian
is zero, $U_F=I$.
In this subsection,
we shall depart from the ideal model \eqref{hamn}-\eqref{bbn}
by adding disorder and perturbations,
\footnote{
  We have also studied different topological Floquet models,
  which are translationally invariant, characterized by the
  non-zero 3d winding number topological invariant
  (and hence non-zero quantized magnetization),
  and having non-zero Floquet Hamiltonian $H_F\neq 0$.
  The results will be reported elsewhere.
}
\begin{align}
  \label{hama}
  H(A)
  &=H_0(t,A)
    \nonumber \\
  &\quad
    +\sum_r w_r c_r^\dag c_r
    + \lambda\sum_r(-1)^{\eta_r} c_r^\dag c^{\ }_r \ ,
\end{align}
where $H_0(t,A)$ is the Hamiltonian introduced in \eqref{hamn}-\eqref{bbn}
coupled to gauge field $A_i$,
the second term is a disorder potential,
where $w_r$ are uncorrelated and can take values between $[-W,W]$ with equal
probability,
and finally, the third term is a clean potential,
where $\eta_r=0$ or $1$ depending on whether $r$ lies in sublattice $A$ or $B$,
respectively.
(Note that $H_0(t,A)$ is zero
for $4T/5 < t < T$ while
the last two terms in \eqref{hama}
are present for all $t$.)
In the following, we shall probe numerically the stability of the response
introduced above.
The disorder term, when sufficiently strong,
guarantees localization.
On the other hand,
what the small $\lambda$ perturbation is expected to do
is to induce a finite bandwidth in the quasi energy spectrum,
and non-zero Floquet Hamiltonian, $H_F\neq 0$:
it can compete with the disorder term.
Both of these terms, when sufficiently strong, can drive the system
away from the topological phase with non-zero $\Theta$ by
going through a continuous transition.
While such transition is interesting,
in this paper, we limit our attention to
small perturbations to the ideal chiral Floquet drive,
and postpone the detailed study of the putative transition
to future works.

\begin{figure}[t]
  \begin{centering}
    \includegraphics[scale=0.65]{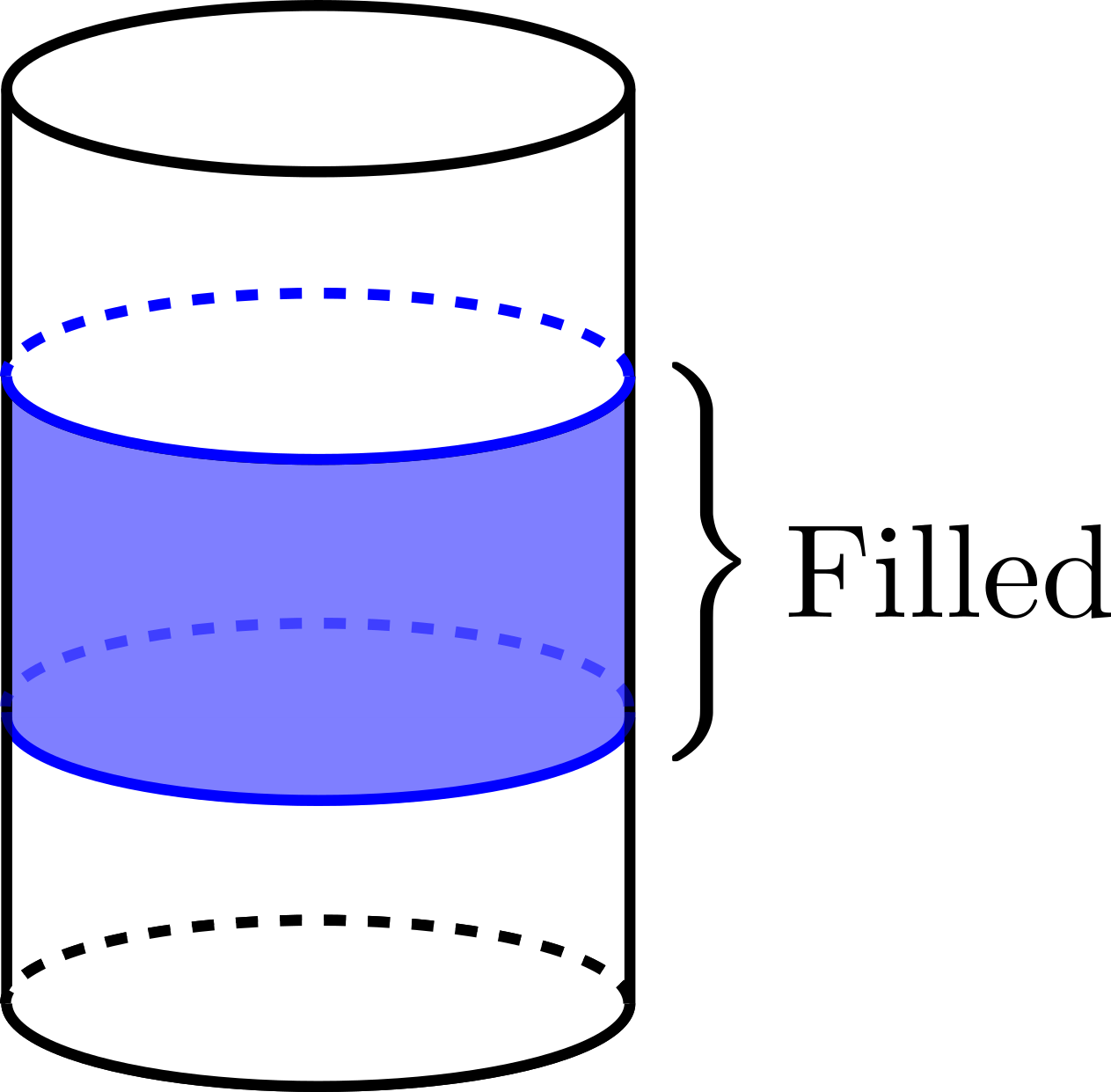}
  \end{centering}
\caption{Representation of the initial state on the cylinder. Site are populated within a strip of height and distance from the boundaries which are longer than the localization length.
\label{strip}
}
\end{figure}

We study the dependence of $\Theta(\alpha)$ on the disorder
strength $W$.
To this aim, we simulated the Hamiltonian \eqref{hama} on a cylindrical lattice of size $L_x=20$ and $L_y=40$.
As initial state, we populated a cylindrical strip of width 16, so that the distance from the
boundaries is sufficiently large compared to the localization length,
see Fig.\ \ref{strip}.
This ensures that we can neglect boundary effects.
The generating functional and the theta angle $\Theta(\alpha)$
are obtained by taking the average of the disorder realizations of the Schwinger-Keldysh trace:
\begin{align}
  e^{iW[A_1, A_2]}=
  \overline{
  \Tr\left[U(W,A_1)\rho_0 U^\dag(W,A_2)\right]}\,
\end{align}
where $\overline{\cdots}$ represents disorder averaging,
in the presence of a fixed background field configuration with
$\int dA_1=2\pi$ and $\int dA_2=0$.
In our simulation, we performed 20 disorder realizations.
We emphasize that, as mentioned below eq.\ (\ref{zaa}),
we need to evaluate $W[A_1,A_2]$ over a very long time in order to isolate the
topological terms.
One can indeed verify numerically that evaluating $W[A_1,A_2]$ over a time which is comparable to the microscopic time scales of the system, $\Theta(\alpha)$ quickly deviates from the unperturbed value as one increases disorder, even if the system is localized.\footnote{Evaluating $W$ over short times is equivalent to having backgrounds $A_{1i},A_{2i}$ that vary fast in time, thus probing quasi-energy scales that are characteristic of the microscopic system.}

\begin{figure}[t]
  \begin{centering}
    (a)\hphantom{AAAAAAAAAAAAAAAAAAAAAAAAAAAAAAA}\\
    \includegraphics[scale=0.55]{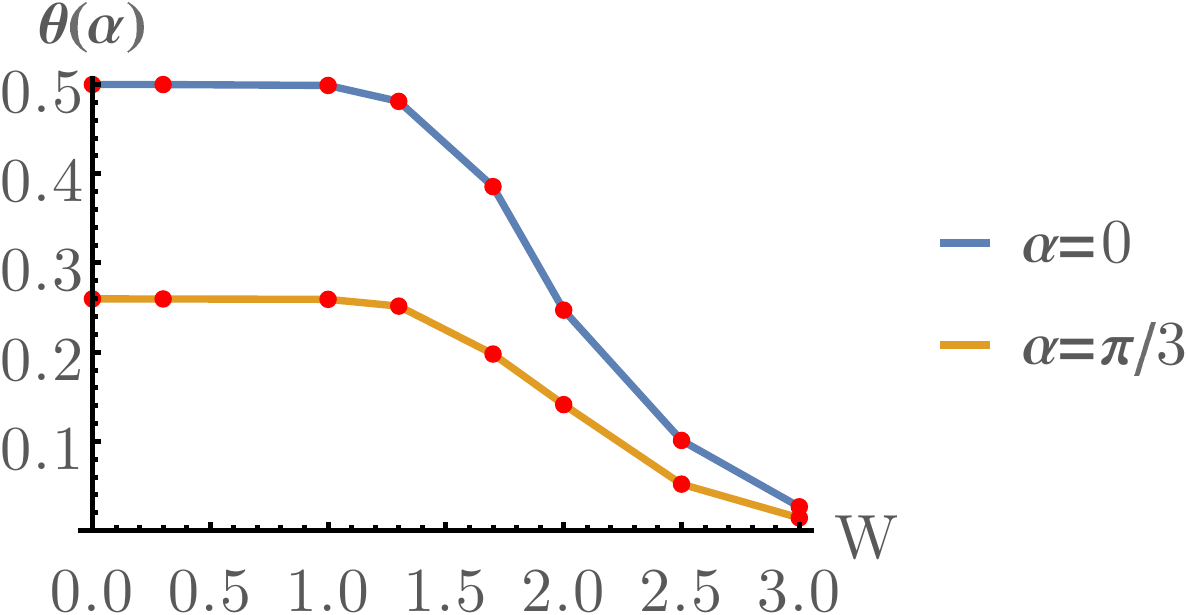}\\
    (b)\hphantom{AAAAAAAAAAAAAAAAAAAAAAAAAAAAAAA}\\
    \includegraphics[scale=0.55]{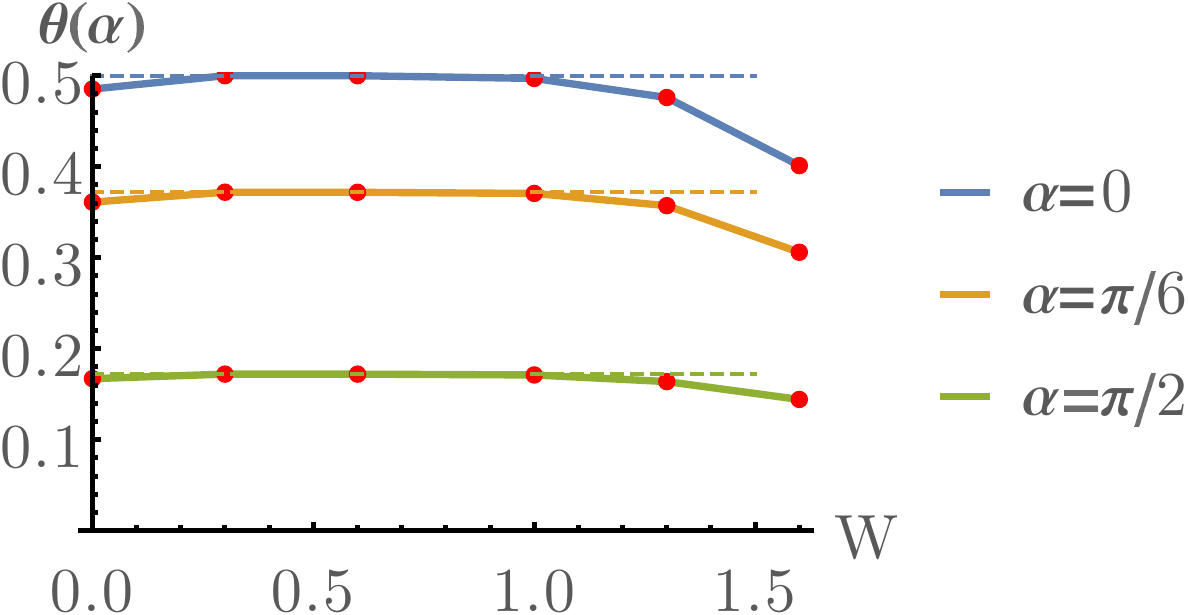}
  \end{centering}
  \caption{
    Plot of $\Theta(\alpha)$ as a function of disorder
    strength $W$, for various values of $\alpha$
    when $\lambda=0$ (a) and $\lambda=0.1 (b)$.
    \label{talpha}
  }
\end{figure}

\begin{figure*}[t]
  \begin{centering}
    \includegraphics[scale=0.45]{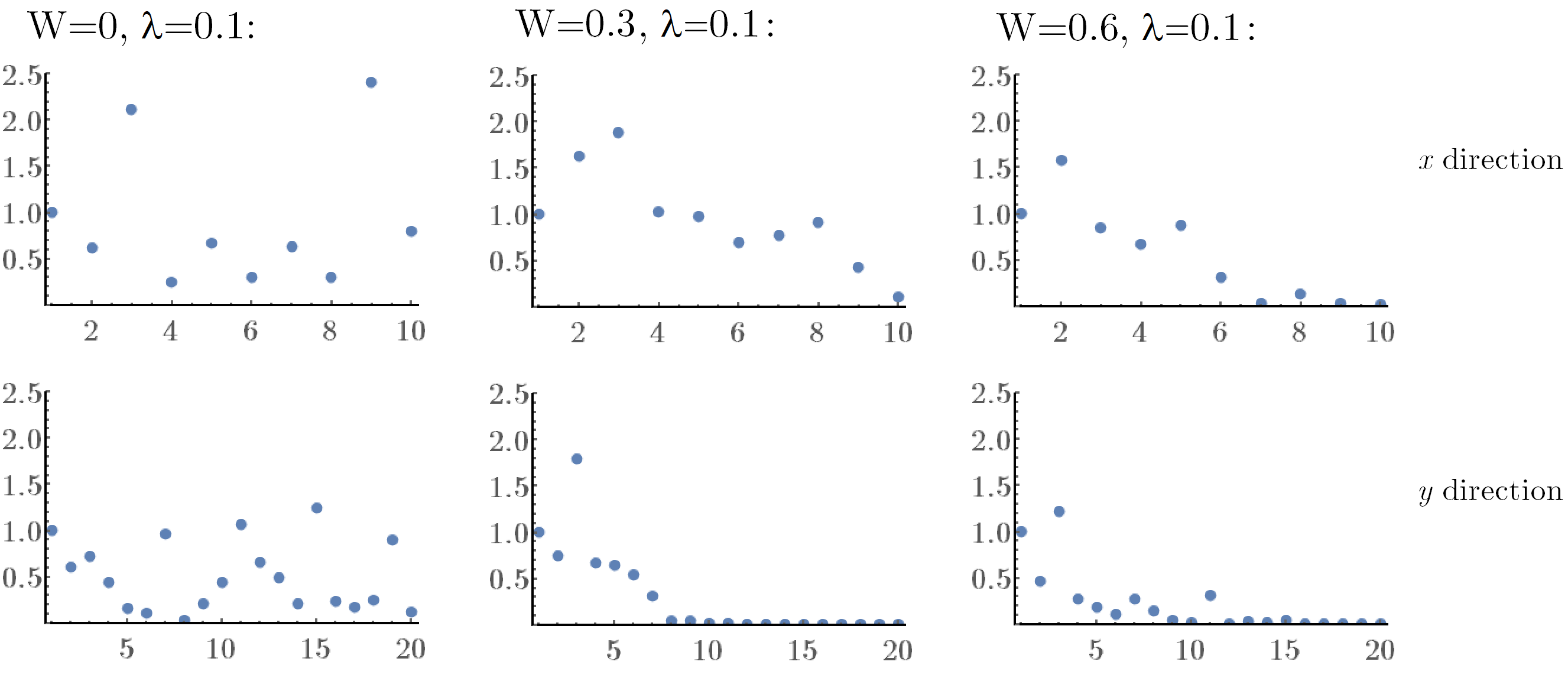}
  \end{centering}
\caption{The plots show $g(r)$ as a function of $r=(x,0)$ in the first row, and $r=(0,y)$ in the second row. As $W$ becomes larger than $\lambda$, correlations between sites drop exponentially.
\label{localiz}
}
\end{figure*}

We first set $\lambda=0$.
Figure \ref{talpha}(a) shows the dependence of $\Theta(\alpha)$ on $W$ for
different values of $\alpha$.
For values of $W$ that are not too large compared to the quasi-energy gap of
$H_0$,
$\vep=2\pi/T=0.8$, one sees the presence of a plateau.
At larger values, the disorder seems sufficiently strong to generate a transition to a topologically trivial state. Confirming this requires more accurate numerical simulations, which we leave for future work.
As a diagnostics of localization, we considered the quantity
\begin{align}
  &g(r)=
  \max_{w_r}\lim_{ {\kappa}\to \infty}
  \left|\frac{\langle r|\, \mathcal{U}({\kappa}T,-\kappa T;A=0)\, |r_0\rangle}
  {\langle r_0|\, \mathcal{U}({\kappa}T,-\kappa T;A=0)\, |r_0\rangle}\right|\ ,
\end{align}
which measures the correlation between a site in the middle of the strip,
$r_0=(0,L_y/2)$ and site $r$
after a long time evolution, and the correlation is maximized over disorder
realizations.
As plotted in Fig.\ \ref{localiz},
we see that the system is localized for all values of $W$ near the plateau


Let us now switch on the third term, the clean potential term.
Figure \ref{localiz} shows that the localized regime holds for $\lambda\ll W$,
as expected.
For $W$ comparable or smaller than $\lambda$, localization is lost and we thus
expect to see a deviation of $\Theta(\alpha)$ from the unperturbed value.
This indeed happens for $W<0.2-0.3$, as shown in Fig.\ \ref{talpha}(b),
consistently with the delocalization-localization transition which happens
around $W=0.2$,
as shown in Fig.\ \ref{localiz}.
As $W$ is increased, localization becomes stronger and $\Theta(\alpha)$ is
brought back to the unperturbed value.
For strong enough disorder, we again see that $\Theta(\alpha)$ drops to zero.

\subsection{Topological chiral Floquet $p,q$ drives}

In this section we apply our response theory to a generalization of chiral
Floquet models which is motivated by and
related to a class of models introduced in
\cite{2017arXiv170307360F,2017PhRvL.118k5301H}.
Those authors found that a class of Floquet systems in two dimensions admits a
topological classification
by a rational number
(GNVW or chiral unitary index),
and characterizes asymmetric quantum information flow at their boundaries.
The topological index can be defined without referencing any symmetry,
and hence these topological Floquet drives do not require
any symmetry for their existence.
From the perspective of the Schwinger-Keldysh effective
field theory approach we are pursuing,
one possible way to detect such topological Floquet drives is to
introduce a ``gravitational'' background,
and look for
a topological term in the gravitational effective action.
Here, in this subsection,
we instead consider a topological Floquet drive with $U(1)$ symmetry,
consisting of multiple species with different charges,
which perform clockwise or counter clockwise chiral motions.
We will see that
the chiral unitary index
is captured by the response
we introduced in the earlier part of this Section,
once we assign charges properly and study the $\alpha$-dependence
of the effective functional.

We start again with a square lattice partitioned into two sublattices,
precisely as described above Eq.\ \eqref{hamn}.
For each site we will now consider a Hilbert space
$\mathcal H_{\mathsf{p}}
\otimes
\mathcal{H}_{\mathsf{q}}$,
where we further factorize
$\mathcal H_{\mathsf{p}}=\bigotimes_{i=1}^{\mathsf{r}} \mathcal H_{\mathsf{p}_i}$
and
$\mathcal H_{\mathsf{q}}=\bigotimes_{i=1}^{\mathsf{s}} \mathcal H_{\mathsf{q}_i}$,
where $\mathsf{p}_i,\mathsf{q}_i$
are prime numbers, and $\mathcal H_{\mathsf{k}}$ has dimension $\mathsf{k}$.
For a given site $r$,
we label states in $\mathcal{H}_{\mathsf{k}}$
by their $U(1)$ charge
as $|r, {n}_{\mathsf{k}}\rangle$
where
${n}_{\mathsf{k}}= 0,\cdots,\mathsf{k}-1$
is the (particle-hole unsymmetrized) particle number.
We then consider the following four-step Floquet drive
\begin{align}
  \label{p q model}
  U_F&=U_4 U_3 U_2 U_1,
  \nonumber \\
  U_n&=\prod_r
  \big(\prod_{\mathsf{p}_i}U_{n,r}^{(\mathsf{p}_i)}\big)
  \big(\prod_{\mathsf{q}_i}U_{n,r}^{(\mathsf{q}_i)}\big)\ ,
\end{align}
where the action of these unitaries on a state
$|r,n_{\mathsf{p}_i}\rangle$ is
\be
U_{n,r}^{(\mathsf{p}_i)}|r,n_{\mathsf{p}_i}\rangle
=e^{i n_{\mathsf{p}_i}A_{r,r+b_n}}|r+b_n,n_{\mathsf{p}_i}\rangle \ ,
\ee
and similarly, the action on a state
$|r,n_{\mathsf{q}_i}\rangle$ is
\be U_{n,r}^{(\mathsf{q}_i)}
|r,n_{\mathsf{q}_i}\rangle=e^{i n_{\mathsf{q}_i}A_{r,r+b_{5-n}}}
|r+b_{5-n},n_{\mathsf{q}_i}\rangle \ .\ee
In summary, $U$ swaps the location of
$\mathsf{p}$-type particles following
counter-clockwise rotation as in the chiral Floquet model
(\ref{hamn})-(\ref{bbn}), while it swaps the location of
$\mathsf{q}$-type particles
following clockwise rotation.
This type of evolution was introduced in
\cite{2017arXiv170307360F,2017PhRvL.118k5301H}.
In our case, we additionally assign $U(1)$ charges to particles so that our
response can directly capture the topology of those models.
In \cite{2017arXiv170307360F,2017PhRvL.118k5301H},
the topological classification was demonstrated by deformation
arguments, where the deformations involved exchanging subspaces of
$\mathcal{H}_{\mathsf{p}}$ of dimension $\mathsf{p}_i$
with subspaces of $\mathcal H_{\mathsf{q}}$ of dimension $\mathsf{q}_i$
whenever $\mathsf{p}_i=\mathsf{q}_i$.
This leads to a topological classification labeled by the factors
$\mathsf{p}_i$ and
$\mathsf{q}_i$ which are pairwise coprime,
i.e. the classification is labeled by $\mathsf{p}/\mathsf{q}$.
Our assignment of charges has been made so that such deformations.
preserve the $U(1)$ symmetry of our Hamiltonian. We can then hope that the
response functional $W[A_1, A_2]$ will automatically capture
the topology property detected and classified by the chiral unitary index.
This will turn out to be the case, which illustrates how $W$ furnishes a systematic diagnostic tool for topology. It would be interesting to deal directly with the neutral system, coupling it to a metric rather than a $U(1)$ gauge field. We leave this for future work.
(See, however, Sec.\ \ref{Geometric response}
for a possible geometric response of topological chiral Floquet drive.)

Let us now obtain the generating functional.
First, the initial density matrix is
$\rho_0={e^{\alpha Q}}/{\Tr e^{\alpha Q}}$, with $Q$ the total charge,
\be Q=\sum_{r}\big(\sum_{i}\tilde n_{\mathsf{p}_i,r}
+\sum_{j}\tilde n_{\mathsf{q}_j,r}\big)\ ,\ee
where we again used particle-hole symmetrized numbers
$\tilde n_{\mathsf{k}}$, in the sense
that the map
$n_{\mathsf{k}}\to \mathsf{k}-1-n_{\mathsf{k}}$
becomes
$\tilde n_{\mathsf{k}}\to-\tilde n_{\mathsf{k}}$.
One then finds
\be\begin{split}
\Tr e^{\alpha Q}=&\prod_r\prod_{\mathsf{p}_i}
\left(\sum_{k=0}^{\mathsf{p}_i-1}
e^{\left(k-\frac{\mathsf{p}_i-1}2\right)\alpha}\right)\\
&\times
\prod_{\mathsf{q}_j}\left(
\sum_{k=0}^{\mathsf{q}_j-1} e^{\left(k-\frac{\mathsf{q}_j-1}2
\right)\alpha}\right)\ .\end{split}\ee
Repeating similar steps as those in the beginning of Sec. \ref{Response functional on the torus}, we obtain the generating functional
\begin{align}
  \label{discw}
  &e^{iW[A_a]}=\frac 1{\Tr e^{\alpha Q}}
    \prod_r
  \nonumber \\
  &\quad
    \times \prod_{\mathsf{p}_i}
\left(\sum_{k=0}^{\mathsf{p}_i-1}
e^{\left(k-\frac{\mathsf{p}_i-1}2\right)\alpha}e^{+ik\int \frac{dt}TB_r}\right)
    \nonumber \\
  &\quad
    \times
    \prod_{\mathsf{q}_i}
\left(\sum_{k=0}^{\mathsf{q}_i-1}
e^{\left(k-\frac{\mathsf{q}_i-1}2\right)\alpha}e^{-ik\int \frac{dt}T B_r}\right)\ .
\end{align}
The structure of this generating functional is similar to that of
\eqref{res1}, where, at each site $r$, we sum over all possible particle numbers and the corresponding flux collected through the micromotion of each particle around the corresponding plaquette. The continuum limit gives
\be
W[A_a]
=\frac{\Theta_{\mathsf{p},\mathsf{q}}(\alpha)}{2\pi}
\int \frac{dt}T\int d^2 r B_a(r)\ ,\ee
where
\be \Theta_{\mathsf{p},\mathsf{q}}(\alpha)
=\theta_{\mathsf{p},\mathsf{q}}+
f_{\mathsf{p},\mathsf{q}}(\alpha)\ ,\ee
with
\begin{align}
  \theta_{\mathsf{p},\mathsf{q}}
  &=\pi\sum_{i=1}^{\mathsf{r}}(\mathsf{p}_i-1)
    -\pi\sum_{i=1}^{\mathsf{s}}(\mathsf{q}_i-1),
    \nonumber \\
  f_{\mathsf{p},\mathsf{q}}(\alpha)
  &=
    \pi\sum_{i=1}^{\mathsf{r}}
    \left(\mathsf{p}_i\coth\frac{\mathsf{p}_i\alpha}2
    -\coth\frac \alpha 2\right)
    \nonumber \\
  &\quad
    -\pi\sum_{i=1}^{\mathsf{s}}
    \left(\mathsf{q}_i\coth\frac{\mathsf{q}_i\alpha}2-\coth\frac \alpha 2\right).
\end{align}
Notice that if there are common factors $\mathsf{p}_i=\mathsf{q}_i$,
the corresponding terms will cancel out in $\Theta_{\mathsf{p},\mathsf{q}}(\alpha)$,
so the continuum limit depends only on factors of the two respective sets
$\{\mathsf{p}_i,\ i=1,\dots,\mathsf{r}\}$
and $\{\mathsf{q}_i,\ i=1,\dots,\mathsf{s}\}$ which are different from
each other, i.e. the response exactly depends on $\mathsf{p}/\mathsf{q}$!
This is fully consistent with the chiral unitary index,
which we now recover as a topological response.
Interestingly, one can see that the phase of \eqref{discw}
is also only dependent on $\mathsf{p}/\mathsf{q}$.
Following the argument of
Sec.\ \ref{Stability under disorder and perturbations},
one then concludes that $\Theta_{\mathsf{p},\mathsf{q}}(\alpha)$ is independent of localization-preserving deformations of the system.


\section{More on effective theory of response}\label{sec:eft}
\label{More on effective theory of response}\label{sec:eft}

In Sec.\ \ref{Chiral Floquet drive},
our primary focus was to derive/calculate
the Schwinger-Keldysh effective response functional
starting from microscopic models such as the 2d chiral Floquet drive.
However, one of the advantages of
the effective field theory approach
is that, based on a few basic principles,
one can put constraints on
allowed terms in the effective action, and
systematically enumerate them,
even without knowing microscopic details of the system.
In this Section, we illustrate the advantage of the effective theory approach to
response by describing two new types of quantized response. We should emphasize
that, while the examples below are consistent with the effective theory of
response, we do not yet know whether and how they can be realized
microscopically, which we leave to future work.

\subsection{Geometric response}
\label{Geometric response}

For the first example, we consider the response to particular geometric deformations. Recall that Floquet systems are invariant under discrete time translation by a period $T$ and that, since we probe the long time behavior, time translation can be viewed as a continuous symmetry. We now gauge this symmetry and introduce a corresponding gauge field. The gauge symmetry acts on the time coordinate as follows
\be
t \rightarrow t + f(\vec r)\,.
\ee
The corresponding gauge field, which we denote as $a_i$, transforms as an abelian gauge field $\delta a_i = -\p_i f(\vec r)$. The gauge invariant generating functional is
\be e^{iW}=\Tr[U(\infty,-\infty;a_{1i})\rho_0 U^\dag(\infty,-\infty;a_{2i})]\ ,
\ee
where $\rho_0$, up to normalization, is the identity, or the projector on a strip such as that in Fig. \ref{strip}. Gauge invariance of the generating functional $W$ implies that the current conjugated to $a_i$ is conserved (in the absence of other external fields):
\be
Q^i = \frac{\delta W}{\delta a_i}\,, \quad \p_i Q^i = 0\,.
\ee
The current $Q^i$ is the (quasi-)energy current since time translation symmetry is responsible for the (quasi-)energy conservation.

To the leading order in derivatives $W$ takes the following form
\be\label{eq:Wenergy}
W=\int \frac{dt}T\int d^2 r\, c_1 (\vep^{ij}\p_i a_{1j}-\vep^{ij}\p_i a_{2j})\ ,
\ee
where as before the time integration is done on $t\in(-{\kappa} T,{\kappa} T)$, ${\kappa}$ a half-integer which we shall take to infinity at the end, and where the factor of $1/T$ has been inserted for convenience. We consider a geometry without boundaries where $a_i$ has a nontrivial flux. The spatial slice is assumed to be flat with the periodic boundary conditions, while $a_i$ is given by
\be
a_i = \omega \epsilon^{ij}r_j\,, \quad \omega = \frac{kT}{L^2}\,,
\ee
where $T$ and $L$ are defined through the twisted spacetime boundary conditions
\begin{gather}\label{iden}
t\sim t+T,\quad r_1\sim r_1+L \ , \quad
\\
(t,r_1,r_2)\sim(t-\omega L (r_1-r_2),r_1,r_2+L)\,.
\end{gather}
Consistency of the above coordinate identifications implies that $k$ is an integer.\footnote{Indeed, the composition of the second identification followed by the third one in (\ref{iden}) results in an identification which is equal to the composition of the third followed by the second one, up to a shift of time $t\sim t-\omega L^2$.}
The flux of $a_i$ will then be $2\omega$ which is quantized. The fact that real time is periodic means that we can consistently place on this geometry only systems whose evolution is truly periodic, $U(t,t_0)=U(t+T,t_0)$.

An example of such system is the unperturbed chiral Floquet model of Sec.\ \ref{Chiral Floquet drive}. Suppose that the system has time-reversal invariance, in the sense that $H^T(t,a_i)=H(-t,-a_i)$.\footnote{One could define a slightly more general notion of time-reversal invariance, i.e. $H^T(t,a_i)=H(t_0-t,-a_i)$. This definition is equivalent to the one in the main text up to translating the definition of the Hamiltonian $H(t,a_i)\to H'(t,a_i)\equiv H(t+t_0/2,a_i)$.} Following the reasoning around \eqref{phcon}, one then requires
\begin{gather} c_1 \int_{-\kappa T}^{\kappa T}\frac{dt}T\int d^2 r\vep^{ij}\p_i
  a_j=c_12\kappa L^22\omega=2\pi
  \nonumber\\
\Longrightarrow\quad c_1\in \frac{\pi}T\mathbb Z\,,
\end{gather}
thus leading to quantization of $c_1$.

Next we discuss the physics interpretation of $c_1$. This coefficient describes
the time-averaged ``thermodynamic'' quantity known as energy magnetization
\cite{2011PhRvL.107w6601Q,2012PhRvL.108b6802N,gromov2015thermal}.
It is defined as the variational derivative\footnote{Here we assume that $W$ depends on $a_i$ only through its flux.}
\be\label{eq:em}
m_E =\frac{\delta W}{\delta (\epsilon^{ij}\p_ia_j)}\,,
\ee
giving
\be\label{eq:defem}
Q^i= \vep^{ij}\p_j m_E\,,
\ee
which justifies the definition. 
With this definition at hand we find that the energy magnetization takes the form
\be
m_E = \frac{c_1}{2\pi T}\,.
\ee

The coefficients $\Theta$ in (\ref{theta}) and $c_1$ are completely independent of each other and provide two independent invariants characterizing a topological Floquet phase.
Comparing to the quantization of the magnetization we have
a relative factor of $ 2\pi/T$.\footnote{This comes from that the ``charge'' of the system with respect to large time translations is $T$, due to the first identification in (\ref{iden}), while in the magnetization case, the $U(1)$ charge is $2\pi$.}

\subsection{Time-ordering sensitive topology}
We now turn to the second extension of our effective response. So far we have seen response of factorized form $W[A_1,A_2]=W_0[A_1]-W_0[A_2]$, i.e. the two Schwinger-Keldysh copies of the background are decoupled, and setting one of them to zero would yield equivalent amount of information. We now show that, at least from the point of view of the effective theory, this is not always the case. The fact that the two copies can talk to each other gives rise to an additional type of topological terms which are related to time ordering.  The most immediate example is the response to a $U(1)$ gauge field in 6+1 dimensions. At leading derivative order, the most general generating functional is
\begin{align}
  \label{WBB}
  W=&\frac 1{4\pi^2}\int \frac{dt}T \int d^6 r\,
      \vep^{ijklpq}
      \Big[3 c_2\p_iA_{rj}\p_k A_{rl}\p_p A_{aq}
  \nonumber \\
    &+(c_3+c_2/4)\p_i A_{aj}\p_k A_{al}\p_p A_{aq}
      \Big]\ ,
\end{align}
where $c_2,c_3$ are constants, and we set the chemical potential to zero for simplicity. Moreover, we conveniently introduced
\be A_{ri}=\frac 12 (A_{1i}+A_{2i}),\quad A_{ai}=A_{1i}-A_{2i}\ .\ee
The part proportional to $c_2$ can be factorized into
\be
c_2 \vep^{ijklpq}
\big(
\p_i A_{1j}\p_k A_{1l}\p_p A_{1q}-\p_i A_{2j}\p_k A_{2l}\p_p
A_{2q}
\big)\ ,
\ee
where the two copies of the background are decoupled as before. This means that, if $c_3=0$, $c_2$ captures information related to the time average of a time-ordered correlation function. The coefficient $c_3$ couples nontrivially $A_{1i}$ and $A_{2i}$, and is related to the time average of a non-time ordered correlation function. To see this more explicitly, let us specialize to the background configuration
\begin{align}
  (A_{s1}, A_{s2})= \frac{B_{s,12}}{2}(-r_2, r_1),
  \nonumber \\
  (A_{s3}, A_{s4})= \frac{B_{s,34}}{2}(-r_4, r_3),
  \nonumber \\
  (A_{s5}, A_{s6})= \frac{B_{s,56}}{2}(-r_6, r_5),
\end{align}
where $s=1,2$ labels the Schwinger-Keldysh copies. Then (\ref{WBB}) gives
\begin{align} \label{bbb0}&\left.\frac{\p^3 e^{iW}}{\p B_{1,12}\p B_{1,34}\p B_{1,56}}\right|_{B=0}=\frac{3\kappa L^6}{2\pi^2}(c_2+2c_3)\ ,
\end{align}
where $L^6$ is the volume of the system, and $\kappa=\frac 12\int\frac{dt}T$ is
a half-integer as usual. Now introduce time-dependent $B_{1,12}(t),B_{1,34}(t),B_{1,56}(t)$.
Using (\ref{tor}),
\begin{align}
  \label{bbb1}
  &
    \left.\frac{\delta^3 e^{iW}}{\delta B_{1,12}(t_1)\delta B_{1,34}(t_2)\delta B_{1,56}(t_3)}\right|_{B=0}
  \nonumber \\
  &\quad
    =\Tr\left[\rho_0
  \mathrm{T}\left(M_{12}(t_1)M_{34}(t_2)M_{56}(t_3)\right)\right]\ ,
\end{align}
where $M_{12}(t)$ is the magnetization operator coupled to $B_{12}$ in the Heisenberg picture,\footnote{For simplicity of illustration, in eq. (\ref{bbb1}) we neglected terms containing higher derivatives of the Hamiltonian with respect to magnetic field, e.g. $\frac{\p^2 H(t,B)}{\p B_{12}\p B_{34}}$. If the Hamiltonian has appreciable nonlinear dependence on the magnetic field, the contribution of such terms in (\ref{bbb1}) may become important.}
\be M_{12}(t)=-U^\dag(t,-\infty)\left(\frac{\p H(t)}{\p B_{12}}\right)_t U(t,-\infty)\ ,\ee
and similarly for $M_{34}(t)$ and $M_{56}(t)$. Note that (\ref{bbb0}) is the time-integrated counterpart of (\ref{bbb1}), so that
\begin{align}
  &c_2+2c_3=\frac{2\pi^2}{3\kappa L^6}
 \int dt_1 dt_2 dt_3
  \nonumber \\
  &\quad
    \times \Tr\left[\rho_0 \mathrm{T}\left(M_{12}(t_1)M_{34}(t_2)M_{56}(t_3)\right)\right]\ ,
\end{align}
i.e. $c_2+2c_3$ is the time integral of a time-ordered 3-point function of magnetization operators. Similarly, one gets
\begin{align} \label{bbb2}&\left.\frac{\p^3 e^{iW}}{\p B_{1,12}\p B_{1,34}\p B_{2,56}}\right|_{B=0}=-\frac{3\kappa L^6}{2\pi^2}c_3\ .
\end{align}
and, using (\ref{tor}),
\begin{align}
  \label{bbb1}
  &
    \left.\frac{\delta^3 e^{iW}}{\delta B_{1,12}(t_1)\delta B_{1,34}(t_2)\delta B_{2,56}(t_3)}\right|_{B=0}
  \nonumber \\
  & \quad
  =\Tr\left[\rho_0 \mathrm{T}\left(M_{12}(t_1)M_{34}(t_2)\right)M_{56}(t_3)\right]\ ,
\end{align}
and we see that $c_3$ is related to the time average of a 3-point function of the same operators as for $c_2+2c_3$, but with different time ordering:
\begin{align}
  &c_3=-\frac{2\pi^2}{3\kappa L^6}
    \int dt_1 dt_2 dt_3
  \nonumber \\
  &\quad
    \times
    \Tr\left[\rho_0 \mathrm{T}\left(M_{12}(t_1)M_{34}(t_2)\right)M_{56}(t_3)\right]\ .
\end{align}

It would be very interesting to realize microscopic systems that lead to such ``time-order sensitive'' topology. We end this section by mentioning that, obviously, one can use standard methods of dimensional reduction to reduce the response (\ref{WBB}) to lower dimensions.

\section{Conclusion}

In this paper, we put forward topological response theory for non-equilibrium topological systems using the Schwinger-Keldysh formalism.
Taking the chiral Floquet drives in two spatial dimensions as an example,
we identify topological terms in the Schwinger-Keldysh generating functional in the presence of static background $U(1)$ gauge field.
As yet another example, in Appendix \ref{Group cohomology models},
we discuss the Schwinger-Keldysh generating functional for topological Floquet unitaries
constructed from group cohomology \cite{2017PhRvB..95s5128R}
with symmetry $G$ in $d$-spatial dimensions.
There again, we identify topological response actions which are elements of
$H^d(G, U(1))$,
in agreement with the previous claim
\cite{2016PhRvB..93x5145V,2016PhRvB..93x5146V,2016PhRvB..93t1103E}.

The presence of these topological terms in the response actions
provides the (many-body) definition of topological Floquet unitaries,
and serves as (many-body) topological invariants.
We expect that the Schwinger-Keldysh effective field theory approach
should work beyond the models studied in this paper,
in generic space dimensions and with various kinds of symmetries.
Nevertheless, the case studied in this paper, namely, the 2d topological
chiral Floquet drive with $U(1)$ symmetry, may be somewhat special in the sense that
the quantized topological term is readily related to
the physically-meaningful response, i.e., quantized magnetization.
For topological terms for other symmetries,
it may be more difficult/non-trivial to relate them to insightful,
physically measurable responses.

Our approach should work even in the absence of symmetry --
one may be able to discuss the coupling of Floquet unitaries
to a background gravitational field.
This may be of particular interest, since there are topological Floquet
unitaries without symmetry
\cite{2017arXiv170307360F,2016PhRvX...6d1070P}.
These systems are characterized by
asymmetric quantum information flow at their boundaries,
and by the quantized edge topological index.
It would be interesting if we can capture the topological index
by properly introducing (a lattice version of) gravitational background
and by the presence of a topological term in the gravitational effective action.
(While we postpone the detailed implementation of this to future works,
we discuss the possible geometric response of the coupling of 2d Floquet drives
in Sec.\ \ref{More on effective theory of response}.)

There are plenty of open questions, such as
an extension of our work to other symmetries,
transitions between different Floquet topological phases,
applications of our formalism to other
non-equilibrium (topological) systems, etc.
Among the most pressing issues is to develop a more comprehensive understanding of
the structure of the Schwinger-Keldysh effective topological action.
For example, we have limited our focus to background field configurations where
$A_{i,1}, A_{i,2}$ are time-independent, and $\alpha$ is a constant.
The motivation for this is that we can exactly compute
the Schwinger-Keldysh effective action
for these choices, but nevertheless, it would be
important to study the effective action
for generic time and for more generic background configurations.

Studying the Schwinger-Keldysh effective action
in the presence of generic background field configurations
seems also important to resolve the following puzzle:
We identified the theta term
in the Schwinger-Keldysh effective topological action for 2d chiral Floquet drives,
which values in $\mathbb{Z}_2$ for closed spatial manifolds and in the presence of particle-hole symmetry.
While the quantization of magnetization can be discussed by using open spatial manifolds,
there is a question if the bulk effective action for closed spatial manifolds
can fully capture the topological nature of 2d chiral Floquet drives.
Also, the theta term is quantized by particle-hole symmetry.
While it does exist in the model we looked at,
one would expect that particle-hole symmetry may be a special property of the
Floquet drive at particular times,
but would ultimately be unnecessary for the fundamental topological property of chiral Floquet drives.

Another point to mention is that
the Schwinger-Keldysh effective topological actions studied
in this paper all have the factorized form, i.e.
the effective response partition function factorizes
between two Schwinger-Keldysh copies.
(See also comments below \eqref{group cohomo result}.)
We may speculate that factorized response partition functions
describe only the subset of topological Floquet drives,
i.e., there may be topological
Floquet drives for which the factorization does not take place,
and the effective functional is given by a complicated polynominal
of $A_a$ and $A_r$.
This may happen in particular in higher dimensions,
as discussed in Sec.\ \ref{More on effective theory of response}.
We leave detailed study of such systems for future works.


{\it Note added:}
While finalizing the manuscript,
\cite{2019arXiv190712228N} appeared on arXiv,
which has some overlap with our work.

\acknowledgements
We thank useful discussion with
Erez Berg, Michael Levin, Yuhan Liu, and Hassan Shapourian.
This work was supported in part by the National Science Foundation grant DMR
1455296 and a Simons Investigator Grant from the Simons Foundation. P. G. was supported by a Leo Kadanoff Fellowship.

\appendix

\section{Group cohomology models}
\label{Group cohomology models}

\label{Group cohomology models}

In this appendix, we consider topological Floquet drives
preserving discrete symmetry $G$.
It has been proposed that topological Floquet systems
in $d$ spatial dimensions
protected by $G$ can be systematically constructed by using the group cohomology
\cite{2016PhRvB..93t1103E,2016PhRvB..93x5145V,2016PhRvB..93x5146V,2016PhRvX...6d1001P}
\begin{align}
  &
  H^{d+1}(G\times \mathbb{Z}, U(1))
  \nonumber \\
  &\quad
    = H^{d+1}(G, U(1))\times H^d (G, U(1)).
    \label{group cohomo classification}
\end{align}
Here, $H^{d+1}(G, U(1))$ corresponds
to static SPT phases in $d$ spatial dimensions
protected by $G$.
On the other hand, $H^d(G,U(1))$
describes non-trivial topological unitaries
specific to Floquet drives.\footnote{
 Recall that static SPT phases described by the group cohomology are expected to
 exhaustive for low spatial dimensions,
 and hence the group cohomology classifies all SPT phases.
}
For example, when $d=1$ and $G=\mathbb{Z}_2\times \mathbb{Z}_2$,
$
H^2(G\times \mathbb{Z}, U(1))
=
H^2(G, U(1))\times H^1 (G, U(1))
=
\mathbb{Z}_2 \times ( \mathbb{Z}_2 \times \mathbb{Z}_2).
$
Here, $H^2(G, U(1))=\mathbb{Z}_2$ corresponds
to the classification of static SPT phases
protected by with $G=\mathbb{Z}_2\times \mathbb{Z}_2$,
which includes the Haldane phase;
$H^1(G,U(1))=\mathbb{Z}_2\times \mathbb{Z}_2$
classifies to non-trivial topological unitaries
specific to Floquet drives.

\subsection{$d=1$ and $G=\mathbb{Z}_2$}

It is also possible to construct explicit lattice models
corresponding to the group cohomology $H^d(G, U(1))$.
As an example, consider the case of $d=1$ and $G=\mathbb{Z}_2$
\cite{2016PhRvB..93x5145V}.
We consider a chain with
the two-dimensional on-site local Hilbert space
spanned by $\{|\pm  \rangle\}$, where $|\pm\rangle$
are the eigen state of the Pauli matrix $\sigma^x$
with eigenvalues $\pm 1$.
The $\mathbb{Z}_2$ symmetry is
generated by
$
  \prod_j \sigma^x_j,
$
where the product is over all sites in the chain.
We consider the Floquet drive of the form:
\begin{align}
  U(t)
&=  e^{-i t H}
  =\prod_j e^{+ i t \sigma^z_j \sigma^z_{j+1}}
    \nonumber \\
  &=
    \prod_j
    \left[
    \cos (t) + i \sigma^z_j \sigma^z_{j+1} \sin(t)
    \right],
\end{align}
where
$H = - \sum_j \sigma^z_j \sigma^z_{j+1}$.
When $t=\pi/2$,
and with PBC,
$U(t)$ is the identity operator,
up to a phase factor:
\begin{align}
  U(\pi/2)
  &=
    \prod_j
    \left[
    i \sigma^z_j \sigma^z_{j+1}
    \right]
    =
    i^{N}
\end{align}
where $N$ is the total number of lattice sites.
While trivial with PBC,
the unitary $U(t=\pi/2)$ is non-trivial
with open boundary condition.

Following the spirits of the preceding sections,
let us now introduce $\mathbb{Z}_2$ gauge fields
$\alpha_{j,j+1}=\pm 1$
for links on the chain, and consider:
\begin{align}
  H[\alpha] &= - \sum_j \sigma^z_j \alpha_{j,j+1}\sigma^z_{j+1}.
\end{align}
Then, our Floquet unitary is:
\begin{align}
  U(t,\alpha)
  &=
  \prod_j e^{+ i t \sigma^z_j \alpha_{j,j+1}\sigma^z_{j+1}}
    \nonumber \\
  &=
    \prod_j
    \left[
    \cos (t) + i \sigma^z_j \alpha_{j,j+1} \sigma^z_{j+1} \sin(t)
    \right].
\end{align}
When $t=\pi/2$ and with PBC,
\begin{align}
  U(t=\pi/2, \alpha)
    &=
    i^N
     \prod_j
     \alpha_{j,j+1}
    =:
    i^N W(\alpha).
\end{align}
Hence, $U(t=\pi/2,\alpha)$ is given by the identity
multiplied by the Wilson loop for $\mathbb{Z}_2$ gauge field
$W(\alpha)=\pm 1$.
It follows that the Schwinger-Keldysh trace
for two Floquet unitaries $U(t,\alpha')$ and $U(t,\alpha)$ is given by
\begin{align}
  Z[t;\alpha,\alpha']
  &=
  \mathrm{Tr}\, \left[
    e^{ -  i t H(\alpha) }
    e^{ +  i t H(\alpha') }
  \right].
\end{align}
In particular, when $t=\pi/2$,
\begin{align}
  Z[t=\pi/2;\alpha,\alpha']
 =
  W(\alpha') W(\alpha).
\end{align}

\subsection{Generic construction}

\paragraph{The Dijkgraaf-Witten theory}
The above construction for $d=1$ and $G=\mathbb{Z}_2$
can be readily extended to more generic cases
\cite{2017PhRvB..95s5128R}.
To describe the generalization,
let us briefly recall the basic ingredients in the Dijkgraaf-Witten theories
\cite{1990CMaPh.129..393D}.
Dijkgraaf and Witten gave a generic construction of
(exponentiated) topological actions $\exp (i I[g, M_n])$
for discrete gauge theories with gauge group $G$,
where $M_n$ is $n$-dimensional Euclidean spacetime, and $\{g\}$
represents a gauge field configuration (see below).
\begin{figure}[t]
\begin{centering}
\hspace{0.8cm}\includegraphics[scale=1]{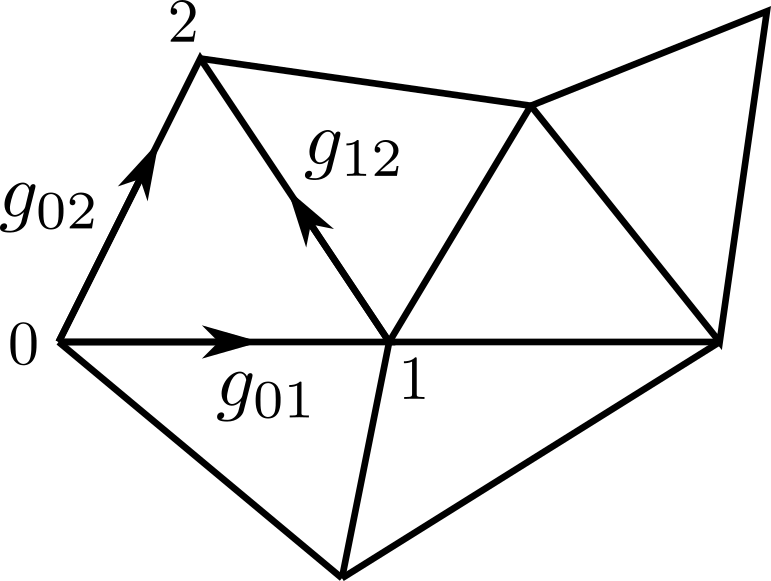}
\end{centering}
\caption{
  Triangulation of spacetime (here for the case of two space time dimensions) for the Dijkgraaf-Witten
  model.
\label{triangles}
}
\end{figure}

The first step of the construction is to triangulate spacetime in terms of
$n$-simplicies (``triangles''),
and assign directions (arrows) to each link.
(E.g., we assign numbers for each vertex in a simplex;
for $i<j$, $\rightarrow$; for $j<i$, $\leftarrow$).
For each elementary triangle ($n$-simplex)
$|\Delta^n|=\pm 1$ represents the orientation of the simplex
with respect to the orientation of spacetime.

We now assign gauge field $g_{ij}\in G$ to each link.
We only consider flat gauge field configurations.
For example, when $n=2$, each triangle
has three links with three gauge fields
$g_{01}$, $g_{12}$, and $g_{02}$;
we impose the flatness condition by
$g_{01}g_{12}=g_{02}$, so that
two out the three gauge fields are independent.
Next, we assign for each $n$-simplex $\Delta^n$
a Boltzmann weight
$\omega_n(g_{01}, g_{12}, g_{23}, \cdots) \in U(1)$.
(For the first entry in $\omega$, we start from
the vertex with no incoming edge, etc.)
Then, the topological action for a given triangulation is given by
\begin{align}
  \exp i I[g,M_n]
  =
\prod_{\Delta^n} \omega_n(\{g\})^{|\Delta^n|}.
  \label{action func dw}
\end{align}

As the final step, we demand the action functional to be
independent of triangulations of $M_n$.
This leads to the condition on $\omega_n$,
the so-called cocycle condition,
which is symbolically given by $d\omega_n=1$.
(We do not write down the definition of $d$ here, but
it can be found in the literature.)
Each solution to this equation gives a topological action
$\exp i I[g,M_n]$.
Inequivalent solutions to the cocycle condition
are classified by the group cohomology $H^n(G, U(1))$.

\paragraph{SPT partition functions}
Equation \eqref{action func dw}
defines the action functional of
the Dijkgraaf-Witten theory (topological gauge theory)
on $M_n$ with gauge group $G$.
Summing over all gauge field configurations $\{g\}$
defines the Dijkgraaf-Witten theory.
On the other hand, the action functional $\exp i I[g, M_n]$ itself
can be viewed as a response theory of an SPT phase protected by symmetry $G$
\cite{2013PhRvB..87o5114C}.

The path integral for the ``matter field'' can be constructed as follows.
We first introduce degrees of freedom living on vertices;
let us call them $v_i \in G$ (where $v_i$ is an element in the group algebra).
We introduce $G$-gauge transformations as
\begin{align}
  v_i \to \alpha_i v_i,
  \quad
  g_{ij} \to \alpha^{\ }_i g_{ij} \alpha^{-1}_j.
\end{align}
Note that combinations $v^{-1}_i g^{\ }_{ij} v^{\ }_j$ are gauge invariant.
In some sense, $\{v_i \}$ can be identified as a gauge transformation.
\begin{widetext}
The Dijkgraaf-Witten action is gauge invariant. Hence, we can write
\begin{align}
  e^{ i I[g, M_n]}
&=
\prod_{\Delta^n}
\omega_n(
v^{-1}_0 g_{01} v^{-1}_1,
v^{-1}_1 g_{12} v^{-1}_2,
v^{-1}_2 g_{23} v^{-1}_3, \cdots)^{|\Delta^n|}
  =
  e^{i S[g,v, M_n]}.
  \label{DW and SPT}
\end{align}
Since $\{v_i\}$ is arbitrary, we can sum over
$\{v_i\}$,
\begin{align}
  Z[g, M_n]
  &=
  e^{ i I[g,M_n]}
  =
  \frac{1}{|G|^{N_v}} \sum_{\{ v_i\}}
    e^{i S[g, v, M_n]}
  \nonumber \\
&=
  \frac{1}{|G|^{N_v}} \sum_{\{ v_i\}}
\prod_{\Delta^n}
\omega(
v^{-1}_0 g_{01} v^{\ }_1,
v^{-1}_1 g_{12} v^{\ }_2,
v^{-1}_2 g_{23} v^{\ }_3,\cdots )^{|\Delta^n|}.
\end{align}
where $N_v$ is the number of vertices.
We can then switch off the background field $g$:
\begin{align}
  Z[M_n]
=
  \frac{1}{|G|^{N_v}}
  \sum_{\{ v_i\}}
\prod_{\Delta^n}
\omega(
v^{-1}_0 v^{\ }_1,
v^{-1}_1 v^{\ }_2,
v^{-1}_2 v^{\ }_3, \cdots)^{|\Delta^n|}.
\end{align}
This can be considered as a partition function of an SPT phase
protected by symmetry $G$.
If there is no boundary on $M_n$, $Z[M_n]=1$.

It is also convenient to introduce
\begin{align}
  \nu(g_0, g_1, g_2, g_3, \cdots)
  \equiv \omega(g^{-1}_0 g^{\ }_1, g^{-1}_1 g^{\ }_2, g^{-1}_2 g^{\ }_3,\cdots).
\end{align}
$\nu$ satisfies (here, we take $n=3$ for simplicity).
\begin{align}
&
 \nu(g g_0, g g_1, g g_2, g g_3) = \nu(g_0, g_1, g_2, g_3),
\nonumber \\
  &
  \nu(g_1, g_2, g_3, g_4) \nu(g_0, g_2, g_3, g_4)^{-1}
  \nu(g_0, g_1, g_3, g_4) \nu(g_0, g_1, g_2, g_4)^{-1}
    \nu(g_0, g_1, g_2, g_3)
  =
    1.
\end{align}
Conversely, when these conditions are satisfied by $\nu$,
one can construct a group cocycle $\omega$ by
\begin{align}
  \omega(g_1, g_2, g_3) = \nu(1, g_1, g_1g_2, g_1 g_2 g_3).
\end{align}
Using $\nu$, the partition function can be written as
\begin{align}
 Z[M_n] =
  \frac{1}{|G|^{N_v}}
  \sum_{\{ v\}}
\prod_{\Delta^n}
\nu(v_0, v_1,v_2, v_3, \cdots)^{|\Delta^n|}.
  \label{SPT part fun}
\end{align}
\end{widetext}

\paragraph{group cohomology models realizing topological floquet drives}
Let us now come back to our question on topological Floquet drives.
Can we construct an explicit unitary operator
with global symmetry $G$ and for a given space dimension $d$,
which, upon introducing a background gauge field, and
then taking the Schwinger-Keldysh trace,
reproduces the response action functional
$\exp i I[g, M_d]$, or more precisely
$\exp i I[g_1,g_2, M_d]$?
We can actually simply take the SPT path integral
\eqref{SPT part fun}
and ``turn'' it into a topological Floquet drive:
Consider a unitary:
\begin{align}
  U(t=T)
  =
  \sum_{\{ v\}}
\prod_{\Delta^d} \nu(v_0, v_1,v_2,\cdots)^{|\Delta^d|}
  | \{v\} \rangle \langle  \{v\} |,
\end{align}
which is completely diagonal.
One can check easily that $U(t=T)$ is the identity operator since
\begin{align}
\langle \{v\}| U(t=T) | \{v'\}\rangle
  &=
\delta_{\{ v\}, \{v'\}}
\prod_{\Delta^d} \nu(v_0, v_1,v_2, \cdots)^{|\Delta^d|}
    \nonumber \\
 &=
    e^{ i S[g=0, v, M_d]} \delta_{\{ v\}, \{v'\}}
    \nonumber \\
 &=
    e^{ i I[g=0, M_d]} \delta_{\{ v\}, \{v'\}},
\end{align}
where we recall \eqref{DW and SPT}.
We can introduce a background gauge field and consider:
\begin{align}
  &
  U(t=T, g)
    \nonumber \\
  \quad
  &=
  \sum_{\{ v\}}
  \prod_{\Delta^d}
  \omega(v^{-1}_0 g_{01} v^{\ }_1, v^{-1}_1 g_{12}v^{\ }_2,\cdots)^{|\Delta^d|}
  | \{v\} \rangle \langle  \{v\} |.
\end{align}
Recalling \eqref{DW and SPT} again,
\begin{align}
\langle \{v\}| U(t=T, g) | \{v'\}\rangle
  &=
    e^{ i S[g, v, M_d]} \delta_{\{ v\}, \{v'\}}
    \nonumber \\
 &=
    e^{ i I[g, M_d]} \delta_{\{ v\}, \{v'\}},
   \label{Group cohomo floquet unitary}
\end{align}
the Schwinger-Keldysh trace
\begin{align}
  Z[t;g_1, g_2] =
  \mathcal{N}^{-1}
  \mathrm{Tr}\,
  \left[
  U(t, g_1)
  U^{\dag}(t, g_2)
  \right]
\end{align}
is given by, when $t=T$,
as a product of the group-cohomology partition functions:
\begin{align}
  Z[t=T;g_1, g_2] =
  \exp\left(+i I[g_1, M_d] -i I[g_2, M_d]\right).
  \label{group cohomo result}
\end{align}

The topological Schwinger-Keldysh response action
\eqref{group cohomo result}
is consistent with the general classification
\eqref{group cohomo classification}
in the sense that the topological term is
\eqref{group cohomo result}
is a member of $H^d(G, U(1))$.
Equation \eqref{group cohomo result}
is also in harmony with \eqref{SK top action chiral floquet}
(although its microscopic counter part \eqref{res1} is more complicated).
We note that in the group cohomology models,
the floquet unitary at $t=T$ is given by
the identity operator, up to an over all phase factor
which is given by $\omega \in H^{d}(G, U(1))$
(see \eqref{Group cohomo floquet unitary}).
As a consequence,
\eqref{group cohomo result} simply
factorizes
$
Z[t=T;g_1, g_2] = \mathcal{N}^{-1}
  \mathrm{Tr}\,
  \left[U(T, g_1)\right]
  \mathrm{Tr}\,
  \left[
  U^{\dag}(T, g_2)
  \right]
$.
This is not the case for the 2d the chiral floquet model;
the floquet unitary at $t=T$ is diagonal
but not proportional to the identity;
\eqref{res1} does not simply factorize.
Nevertheless,
$
  \mathrm{Tr}\,
  \left[U(T, A_1)
  U^{\dag}(T, A_2)
  \right]
$
depends only on the difference,
$A_a = A_1 - A_2$,
and for smooth (long-wave length) configurations,
the topological term can still be written
in the factorized form \eqref{SK top action chiral floquet}.
The factorization of the
response Schwinger-Keldysh action
has an affinity with
the proposed group cohomology classification
\eqref{group cohomo classification},
in which we do not see any inkling of
the Schwinger-Keldysh formalism;
at least naively,
the group cohomology $H^{d+1}(G\times \mathbb{Z}, U(1))$
is expected to classify
the Euclidean path integral
without using the Schwinger-Keldysh copies.
Nevertheless,
the calculations presented here
show the factorization of
the Schwinger-Keldysh action,
and the topological terms in each copies,
$\mathrm{Tr}\, \left[U(t, g_1)\right]$ and
$\mathrm{Tr}\, \left[U^{\dag}(t, g_2) \right]$,
are labeled by $H^d(G, U(1))$;
we thus land on \eqref{group cohomo classification}.

%

\section{Channel-state map approach}
\label{Channel-state map approach}


In this appendix, we introduce an approach based
on the so-called channel-state map
(the Choi-Jamio\l kowski isomorphism),
which maps arbitrary unitary operator,
acting on a Hilbert space $\mathcal{H}$,
to a state living in a bigger (doubled) Hilbert space,
$\mathcal{H}\otimes \mathcal{H}^*$.
In physics context, this has been used
in the thermofield double state,
and the thermo field dynamics
\cite{Umezawa:1993yq}.
This channel-state map allows us
to ``transplant'' the approaches to static (topological) states
to (topological) unitaries and
develop an effective response theory.

As the Schwinger-Keldysh formalism,
the thermo field dynamics provides a framework
to describe the real-time non-equilibrium dynamics of
finite temperature systems.
In particular, at equilibrium, the thermo field dynamics
and the Schwinger-Keldysh formalism are equivalent.
In some sense, the purpose of this section is to
``redo''
what we have achieved in the main text
using the Schwinger-Keldysh formalism
by using the thermo field dynamics (channel-state map),
although
the precise relation between
the Schwinger-Keldysh approach and
the approach presented here is not entirely clear.

As a byproduct of using the channel-state map,
we will be able to make a contact with
the common trick used,
e.g., in Ref.\ \cite{2017PhRvB..96o5118R}
to derive the periodic table of
non-interacting Floquet fermion systems
(the ``Hermitian map'').
There,
one first artificially doubles the original
(single-particle) Hilbert space,
and then embeds Floquet unitaries
into a Hermitian operator (``Hamiltonian'')
acting on the doubled Hilbert space.
We will provide a point of view in terms of the channel-state map.


\subsection{Operator-state map}

The channel-state map
(the Choi-Jamio\l kowski isomorphism)
applies to an arbitrary quantum channel
(trace-preserving completely positive (TPCP) map),
and maps it to a quantum state (density matrix)
in the doubled Hilbert space.
In simplest cases,
it maps a unitary operator $U$
acting on the Hilbert space $\mathcal{H}$
to a (pure) state
in the doubled Hilbert space
$\mathcal{H}\otimes \mathcal{H}^*$:
\begin{align}
|U\rangle\!\rangle = (I \otimes U) |\Omega \rangle\!\rangle
\end{align}
where $|\Omega \rangle\!\rangle$ is a maximally entangled state
\begin{align}
  |\Omega \rangle\!\rangle = ( {1}/\sqrt{\mathcal{N}} )
  \sum_{i} |i\rangle |i\rangle^*,
  \quad
  \mathcal{N}\equiv \mathrm{dim}\, \mathcal{H} = \mathrm{Tr}_{\mathcal{H}}\, I.
\end{align}
Essentially the same mapping from an operator to a state
is used in the context of the thermofield double state, where a thermal density operator is mapped to
a state (thermofield double state) in the doubled Hilbert space.
Observe that the overlap of two states corresponding to unitaries $U$ and $U'$ is
\begin{align}
\langle \!\langle U |U'\rangle\!\rangle
  =
  (1/\mathcal{N})
  \mathrm{Tr}_{\mathcal{H}}
  \left[ U^{\dag} U^{\prime} \right],
\end{align}
which can be represented as a Schwinger-Keldysh
path-integral with
the infinite temperature thermal state
as the initial state.

\subsection{Fermionic chiral floquet drive}

Let us now consider
a fermionic system described by
a set of fermion annihilation/creation operators,
$
\{
\hat{\psi}^{\ }_a, \hat{\psi}^{\dag}_b
\}=\delta_{ab}$.
Here,
$a,b=1,\ldots, N$ and
$N$ is the number of independent ``orbitals'',
i.e., the dimension of the single-particle Hilbert space.
Following our general discussion,
we double the fermion Fock space,
$\mathcal{H}\to \mathcal{H}\otimes \mathcal{H}$,
and consider the state
$
  ( I \otimes \hat{U}) |\Omega \rangle\!\rangle
$
where $|\Omega\rangle\!\rangle$ is a suitable maximally entangled state
in the doubled Hilbert space.
For the current example,
an appropriate choice of $|\Omega \rangle\!\rangle$ is given by
\begin{align}
  |\Omega \rangle\!\rangle \equiv
    \prod_{a}
    \frac{1}{\sqrt{2}}
  \left[
  \hat{\psi}^{\dag}_{aA} + \hat{\psi}^{\dagger}_{aB}
  \right]
 |0\rangle\!\rangle,
\end{align}
where
we now have two independent sets of
fermion annihilation/creation operators,
$\{ \hat{\psi}^{\ }_{aA}, \hat{\psi}^{\dag}_{aA}\}$
and
$\{ \hat{\psi}^{\ }_{aB}, \hat{\psi}^{\dag}_{aB}\}$,
acting on each copy of the fermion Fock space, $\mathcal{H}_A$ and $\mathcal{H}_B$.
Note that for a given ``site'' $a$,
the state is
a equal superposition of
states of charge $q$ on $A$
and $-q$ on $B$,
where $q=\pm 1/2$ is the total particle number measured
from half-filling,
$
  \big[
  \hat{\psi}^{\dag}_{aA} + \hat{\psi}^{\dagger}_{aB}
  \big]
 |0\rangle\!\rangle
 =
 |10\rangle\!\rangle
+
 |01\rangle\!\rangle
=
\sum_{q}
|q+1/2, -q+1/2\rangle\!\rangle
$.
The state dual to $\hat{U}$ can be constructed accordingly
as
$
| U\rangle\!\rangle =
(I\otimes \hat{U})| \Omega \rangle\!\rangle$.

We will be interested in ``short-range correlated states''.
I.e.,
all equal time correlation functions:
$
  \langle \! \langle U|
  \hat{\Psi}^{\dag}_i \cdots \hat{\Psi}^{\ }_j \cdots
  |U\rangle \! \rangle
  =
\langle \! \langle \Omega |
  \hat{U}^{\dag} \hat{\Psi}^{\dag}_i \hat{U} \cdots
\hat{U}^{\dag}\hat{\Psi}^{\ }_j \hat{U} \cdots
  |\Omega \rangle \! \rangle
$
are local in the sense that they decay exponentially in distances.
As far as evolution driven by $U$ is ``local'' or ``non-ergodic'',
as in the case of many-body localized evolution,
we expect that $|U\rangle\!\rangle$ can be
treated as a ground state of a gapped system.
%

The reference state $|\Omega \rangle\!\rangle$
is a unique ground state of the
``parent'' Hamiltonian
\begin{align}
  \hat{K}_0
  =\sum_{a,b}
\big(
  \hat{\psi}^{\dag}_{aA} \hat{\psi}^{\ }_{bB}
  + h.c.
\big),
\end{align}
acting on $\mathcal{H}\otimes \mathcal{H}$.
Similarly, $|U\rangle\!\rangle$
is a unique ground state of
\begin{align}
  \hat{K} &=
  (I\otimes \hat{U})
  \hat{K}_0
  (I\otimes \hat{U}^{\dag})
            \nonumber  \\
   &=\sum_{a,b}
\big(
     \hat{\psi}^{\dag}_{aA}
     \hat{U}\hat{\psi}^{\ }_{bB} \hat{U}^{\dag}
  + h.c.
\big).
\end{align}

Let us have a closer look at of this mapping
for the case of
a quadratic Hamiltonian
$
\hat{H} = \sum_{a,b=1}^{N} \hat{\psi}^{\dag}_a \mathcal{H}_{ab} \hat{\psi}^{\ }_b
$
and the corresponding unitary evolution operator $\hat{U}$.
The many-body unitary operator $\hat{U}$ defines
a unitary matrix $U$ through
$
\hat{U} \hat{\psi}_a \hat{U}^{\dag} = \mathcal{U}_{ab} \hat{\psi}_b.
$
The state
$|U\rangle\!\rangle
=  (I\otimes \hat{U}) |\Omega \rangle\!\rangle
$
can be explicitly calculated easily:
\begin{align}
  |U \rangle\!\rangle
  &=
  \prod_{a}
    \frac{1}{\sqrt{2}}
  \left[
  \hat{\psi}^{\dag}_{aA} + \sum_b \mathcal{U}^{\dag}_{ab} \hat{\psi}^{\dagger}_{bB}
  \right]
 |0\rangle\!\rangle.
\end{align}
The parent Hamiltonian is
\begin{align}
  \hat{K} &
      = \sum_{i,j=1}^{2N} \hat{\Psi}^{\dag}_i \mathcal{K}_{ij} \hat{\Psi}_j
      =
      \sum_{a,b=1}^N
        \left[
        \hat{\psi}^{\dag}_{aA} \mathcal{U}^{\ }_{ab} \hat{\psi}^{\ }_{bB} +
        \hat{\psi}^{\dag}_{bB} \mathcal{U}^{\dag}_{ab} \hat{\psi}^{\ }_{aB}
        \right],
\end{align}
where $\Psi^{\dag}, \Psi$ and
the $2N\times 2N$ matrix $\mathcal{K}$ are given by
\begin{align}
  \hat{\Psi}^{\dag}=\left[
  \begin{array}{cc}
    \hat{\psi}^{\dag}_A & \hat{\psi}^{\dag}_B
  \end{array}
  \right],
                          \,
  \mathcal{K} =
  \left[
  \begin{array}{cc}
    0 & \mathcal{U} \\
    \mathcal{U}^{\dag} & 0
    \end{array}
  \right],
               \,
  \hat{\Psi} =
  \left[
  \begin{array}{c}
    \hat{\psi}_A \\
    \hat{\psi}_B
    \end{array}
  \right].
\end{align}
Passing from the original (single-particle) unitary matrix $\mathcal{U}$ to
the hermitian matrix $\mathcal{K}$
is the ``Hermitian map'' used in, e.g.,
Ref.\ \cite{2017PhRvB..96o5118R}
to derive the periodic table of Floquet topological systems.
While the original Hamiltonian $\mathcal{H}$ is a member of
symmetry class A (if we do not assume any symmetry),
$\mathcal{K}$ is a member of symmetry class AIII:
$\mathcal{K}$ is invariant
under the following antiunitary transformation
(chiral symmetry):
\begin{align}
\hat{S}\, \hat{\psi}^{\ }_{aA} \, \hat{S}^{-1} &= \hat{\psi}^{\dag}_{aA},
                                                 \quad
\hat{S}\, \hat{\psi}^{\ }_{aB} \, \hat{S}^{-1} = - \hat{\psi}^{\dag}_{aB},
                                                 \nonumber \\
\hat{S}\, \hat{\psi}^{\dag}_{aA} \, \hat{S}^{-1} &= \hat{\psi}^{\ }_{aA},
                                                 \quad
\hat{S}\, \hat{\psi}^{\dag}_{aB} \, \hat{S}^{-1} = - \hat{\psi}^{\ }_{aB}.
\end{align}
This transformation can be considered as a
composition of the modular conjugation operator (tilde conjugation operator)
in the Tomita-Takesaki theory (the thermofield double theory),
and the swap operation $\hat{\psi}_A \leftrightarrow \hat{\psi}_B$.


%
%
%


\subsection{Building effective response field theories by dimensional reduction}

Note that the spectrum of $K$ is gappled and completely ``flat'':
Its eigenvalues are all either $\pm 1$.
Any $K$ of this form can be obtained from
a more ``physical'' Hamiltonian preserving chiral symmetry
and having a energy gap by spectral flattening
\cite{2010NJPh...12f5010R}.
As $|U\rangle\!\rangle$ is
realized as a unique ground state of a gapped Hamiltonian
$\hat{K}$, its topological properties can be
studied and classified by using the techniques
of static (symmetry-protected) topological phases.

With the help of the operator-state map,
and assuming the presence of
reasonable parent Hamiltonians $\hat{K}$,
we now proceed to develop effective response field theories.
We henceforth resurrect the so-far neglected time-dependence
in the unitaries, $\hat{U}(t)$,
and work with periodic unitaries, $\hat{U}(t+T)=\hat{U}(t)$.

Following the recipe of deriving effective
response field theories for static topological phases,
we introduce
a imaginary-time spacetime path integral of type \eqref{path integral}.
Naively, this would introduce yet another time than
the real time $t$, which simply enters in the path integral
as a parameter;
For a Floquet system living
on physical $(d+1)$-spacetime dimensions,
we have $(d+2)$-dimensional spacetime.
As we will see,
this issue can be naturally solved
if we make a contact with the theory of adiabatic quantum pump,
a typical example of which is the Thouless pump
in (1+1)-dimensional system.
The topological properties of
Floquet unitary operators in $(d+1)$-dimensions,
may be related to $(d+2)$-dimensional topological phases.
The response field theory of the latter can be
dimensionally reduced to describe the target $(d+1)$-dimensional physics.
This means that
we are effectively considering
the adiabatic evolution of
Floquet unitaries $U(t)$ as a function of $t$,
while the time-evolution of
physical states by Floquet unitaries
are not adiabatic in general.

Observe also that, if we start from systems with no-symmetry (class A),
mapping unitaries to states by the channel-state map
transforms the symmetry class from A to AIII by
working with the doubled Hilbert space.
This is in a perfect harmony with
the above dimensional shift $(d+1) \to (d+2)$,
and
with the Bott periodicity.

%

Now, following the recipe of deriving effective
response field theories for static topological phases,
we introduce a background $U(1)$ gauge field $V=V_{\mu}dx^{\mu}$.
This in principle has nothing to do with
physical electromagnetic $U(1)$ gauge field $A=A_{\mu}dx^{\mu}$,
as it is introduced in the doubled Hilbert space.
(See the comments below, though).
By integrating over the matter field, we would then arrive at
the effective response theory.
Since we are in (3+1)d, and since our
Hamiltonian $K$ belongs to class AIII,
the topological part of the resulting
effective action is given by the axion term:
\begin{align}
  \label{axion term}
  W_{{\rm eff}}[V]
  &=
    \frac{\theta}{8 \pi^2}
  \int du d^3x\,
  \varepsilon^{\mu\nu\kappa\lambda}
    \partial_{\mu}V_{\nu} \partial_{\kappa}V_{\lambda},
\end{align}
where
$u$ is the fictitious imaginary time, which is
analytically continued to the Lorentz signature in Eq.\ \eqref{axion term}.
The $\theta$ angle here is pinned to quantized values,
$\theta = \pi \times ({\rm integer})$,
by the chiral symmetry.

The next step is to
dimensionally reduce
this action to (2+1)d:
we shrink the size of $z$-direction $L_z$ to zero,
and decompose the vector field $V_{\mu}$ in (3+1)d
into vector and scalar fields in (2+1)d.
Explicitly, we introduce the scalar in terms
of the $z$-component of $V$ as
$\Phi(u,x,y) = V_z(u,x,y)/L_z$.
The resulting action is given by
\begin{align}
  W_{{\rm eff}}[V] =
  \frac{\theta}{2\pi^2} \int du dxdy\,
  \varepsilon_{\mu\nu\lambda}
  \partial_{\mu} \Phi\, \partial_{\nu}V_{\lambda}.
\end{align}

From the effective response action we can
read off the topological responses
and also topological invariants.
(See, for example, Ref.\ \cite{2018PhRvB..98c5151S}.)
We consider the magnetic field
${B}_V= \epsilon_{ij}\partial_i V_j$,
and
define the local magnetization density
${M}(u,x,y)$ by
\begin{align}
  {M}(u, x,y)
  \equiv
  \frac{\delta W_{{\rm eff}}}{\delta {B}_V}
  =
  \frac{\theta}{2\pi^2}
  \partial_{u}\Phi(u, x, y).
\end{align}
We also introduce
the magnetization per unit volume
\begin{align}
  {m}(u)
  =
  \frac{1}{ {\it Vol}}
  \int dxdy\,
  M(u, x,y).
\end{align}
Then,
the time-average of $m(u)$ is
\begin{align}
  \overline{m(u)}
  &=
  \frac{1}{T}
  \int^T_0
  du
  \frac{\theta}{2\pi^2}
  \frac{d \Phi}{du}
  =
  \frac{1}{T}
  \frac{\theta}{2\pi^2}
    \left[
    \Phi(T) -\Phi(0)
    \right]
    \nonumber \\
  &=
    \frac{1}{T}
  \frac{\theta}{\pi}.
\end{align}
%
Recalling $\theta = \pi \times {\it integer}$,
this fictitious magnetization is quantized.
Its connection to the physical magnetization is unclear, though.
Nevertheless, we note
that it can be shown that
the $\theta$-angle is given in terms of the winding
number topological invariant associated with the unitary matrix
$\mathcal{U}(t)$ \cite{2010NJPh...12f5010R}.
This then proves, indirectly, that the fictitious magnetization agrees with the physical magnetization discussed
in \cite{2017PhRvL.119r6801N}.

\bibliographystyle{ieeetr}
\bibliography{reference}

\end{document}